\documentclass[showpacs,amsmath,amssymb,aps,pra,longbibliography,10pt,reprint,superscriptaddress,showkeys]{revtex4-1}
\usepackage{graphicx}
\usepackage{bm}
\usepackage[breaklinks=true,colorlinks=true,linkcolor=blue,urlcolor=blue,citecolor=blue]{hyperref}
\usepackage{graphics}
\usepackage{dcolumn}
\usepackage{amsmath,amssymb}
\usepackage{mathdots}
\usepackage{natbib}
\usepackage{soul,color}
\usepackage{epstopdf}
\usepackage[note-name]{notes2bib}

\begin{document}

\title{\large Moving flux quanta cool superconductors by a microwave breath}

\author{O. V.~Dobrovolskiy}
    \email{oleksandr.dobrovolskiy@univie.ac.at}
    \affiliation{Faculty of Physics, University of Vienna, 1090 Vienna, Austria}
    \affiliation{Physics Department, V. Karazin National University, 61077 Kharkiv, Ukraine}
\author{C. Gonzalez-Ruano}
\author{A. Lara}
    \affiliation{Dpto. Fisica de la Materia Condensada C03, IFIMAC and INC,\\ Universidad Aut\/onoma de Madrid, Cantoblanco 28049 Madrid, Spain}
\author{R. Sachser}
    \affiliation{Physikalisches Institut, Goethe University, 60438 Frankfurt am Main, Germany}
\author{V. M. Bevz}
    \affiliation{Physics Department, V. Karazin National University, 61077 Kharkiv, Ukraine}
\author{V. A. Shklovskij}
    \affiliation{Physics Department, V. Karazin National University, 61077 Kharkiv, Ukraine}
\author{A.~I.~Bezuglyj}
    \affiliation{Physics Department, V. Karazin National University, 61077 Kharkiv, Ukraine}
    \affiliation{Institute for Theoretical Physics, NSC-KIPT, Kharkiv, Ukraine}
\author{R.~V.~Vovk}
    \affiliation{Physics Department, V. Karazin National University, 61077 Kharkiv, Ukraine}
\author{M. Huth}
    \affiliation{Physikalisches Institut, Goethe University, 60438 Frankfurt am Main, Germany}
\author{F. G. Aliev}
    \affiliation{Dpto. Fisica de la Materia Condensada C03, IFIMAC and INC,\\ Universidad Aut\/onoma de Madrid, Cantoblanco 28049 Madrid, Spain}
\date{\today}

\begin{abstract}
Almost any use of a superconductor implies a nonequilibrium state. Remarkably, the non-equilibrium states induced by a microwave stimulus and the dynamics of magnetic flux quanta (Abrikosov vortices) can give rise to strikingly contrary effects: A sufficiently high-power electromagnetic field of GHz frequency can stimulate superconductivity, whereas fast vortex motion can trigger an instability abruptly quenching the superconducting state. Here, we advance or delay such dynamical quenching of the vortex state in Nb thin films by tuning the power and frequency of the microwave ac stimulus added to a dc bias current. The experimental findings are supported by time-dependent Ginzburg-Landau simulations and they can be explained qualitatively based on a model of ``breathing mobile hot spots'', implying a competition of heating and cooling of quasiparticles along the trajectories of moving fluxons whose core sizes vary in time. In addition, we demonstrate universality of the stimulation effect on the thermodynamic and transport properties of type II superconductors.
\end{abstract}

\maketitle

Superconductors in presence of high-frequency electromagnetic fields are exploited in diverse applications, such as the Josephson voltage standard  \cite{Bar82boo}, THz and GHz radiation emitters  \cite{Wel13nph,Dob18nac}, photon detectors  \cite{Mar13nph}, as well as circuits for quantum electrodynamics  \cite{Gux17phr} and quantum computing  \cite{Dev13sci}. Furthermore, the interplay of Meissner screening and magnetic flux with spin-wave dynamics at microwave frequencies has recently become a matter of intensive research in the rapidly developing domains of superconducting spintronics  \cite{Lin15nph,Kim18prl,Jeo18nam,Jeo19pra} and magnon fluxonics \cite{Gol18afm,Dob19nph}. This interest is largely because of the interacting superconducting and ferromagnetic orders \cite{Wan06nat,Bes14prb,Sto18adv}, and the high sensitivity of microwave techniques involving superconductors \cite{Emb17nac,Yue17apl,Gol18jap} for probing the quasiparticle and spin dynamics \cite{Kim18prl,Jeo18nam,Jeo19pra} as well as for control and readout of qubits \cite{Gir14inp}. The rich physics of interaction of superconductors with a high-frequency excitation involves many complex mechanisms related to the dynamics of nonequilibrium quasiparticles \cite{Lar75etp,Lar86inb,Bez92pcs,Gra81boo,Kop01boo,Ser18prl}, the superconducting gap evolution \cite{Eli70etp,Gur08prb}, and the dynamics of Abrikosov vortices \cite{Pom08prb,Kog18prb}.

Remarkably, the non-equilibrium states in superconductors generated by a microwave stimulus and the Abrikosov vortex dynamics can imply conceptually opposite effects. Thus, while superconductivity is known to be destroyed by high temperatures, currents and magnetic fields, a sufficiently high-power electromagnetic field of subgap frequency may stimulate superconductivity itself \cite{Wya66prl}. This counterintuitive effect was explained as a consequence of an irradiation-induced redistribution of quasiparticles away from the superconducting gap edge \cite{Eli70etp}. While microwave-stimulated superconductivity was since then observed in various systems \cite{Zol13ltp,Bec13prl,Vis14prl}, no theory of this effect in the \emph{presence of vortices} is available so far. At the same time, investigations of stimulated superconductivity in the vortex state could have important implications both inside and outside the condensed matter physics community, as mapping solutions of astrophysics problems to scalar condensates allows for modeling the physics of black holes and gravity upon holographic superconductors \cite{Har08prl}. While the latter can exhibit vortices \cite{Mon09prl} and vortex-flow states \cite{Mae11prd}, the possibility of stimulation of holographic superconductors is under current debate \cite{Bao11hep,Nat13hep}.

The presence and motion of vortices is expected to modify the signatures of microwave-stimulated superconductivity due to the variation of the order parameter in space and time, the friction-induced heating of quasiparticles in the vortex cores \cite{Cle68prl,She11prl,Gur08prb}, and the energy leaking outside of the core at large vortex velocities \cite{Lar75etp,Lar86inb,Bez92pcs}. The latter results in a vortex-core shrinkage, a further acceleration of the flux flow, and the associated quench of the low-resistive state due to the flux-flow instability \cite{Lar75etp,Lar86inb,Bez92pcs}. While superconductivity in the presence of a high-frequency excitation and non-equilibrium effects related to vortex dynamics have recently received much attention theoretically \cite{Sem16prl,Tik18prb,Kog18prb,Shk17prb,Vod17pra,Yan18prb} and experimentally \cite{Leo11prb,Sil12njp,Per05prb,Wor12prb,Che14apl,Lar15nsr,Lar17pra,Mad18sca,Dob19pra,Bez19prb,Dob19rrl}, the microwave-stimulated and the vortex-dynamics-generated nonequilibrium states have never been studied simultaneously so far.
\begin{figure*}[tbh!]
    \centering
    \includegraphics[width=0.86\linewidth]{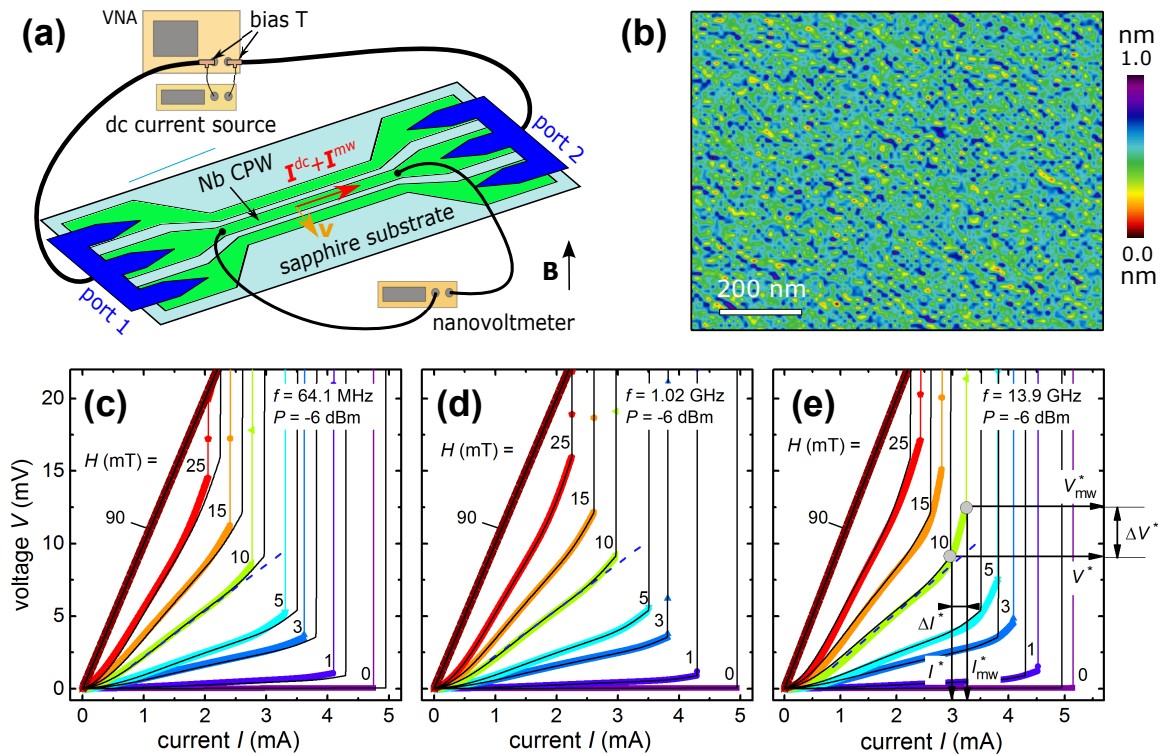}
    \caption{
    \textbf{Microwave current control of dc-driven quenching of superconductivity}.
    (a) Experimental geometry. The Nb coplanar waveguide is in a perpendicular magnetic field with induction $\mathbf{B}$.
    The joint action of direct and microwave currents $I^{\mathrm{dc}} + I^{\textrm{mw}}$ exerts a Lorentz force on the vortices and makes them move with velocity $\mathbf{v}$ across the central conductor of the waveguide. The voltage drop along the central conductor allows one to distinguish between the different resistive states of the sample.
    (b) Atomic force microscopy image of the surface of the Nb film.
    (c-e) The current-voltage ($I$-$V$) curves of the Nb film at a temperature $T = 0.988 T_\mathrm{c}$ for a series of magnetic field values, as indicated,     in the unexcited state (solid lines) and in the presence of an ac current (symbols).
    Depending on the ac frequency at the fixed ac power level $-6$\,dBm, the range of dc currents relating to the nonlinear resistive regime
    shrinks (c), remains the same (d) and expands (e) as compared to the non-excited regime.
    The definition of the instability current $I^\ast$ and the instability voltage $V^\ast$ at the last point before the jump to the normal state is indicated in panel (e). $I^\ast_\mathrm{mw}$ and $V^\ast_\mathrm{mw}$ designate the values deduced from the $I$-$V$ curves in the presence of a microwave ac current.
    The dashed straight lines in panels (c) to (e) are guides to the eye.
    }
    \vspace{2mm}
    \label{fCVC}
\end{figure*}

\vspace{3mm}
Here, we investigate the competition between quenching and stimulation of superconductivity in the GHz-frequency (ac+dc)-driven nonlinear resistive regime in Nb thin films. Under optimized excitation conditions we demonstrate a pronounced extension of the low-dissipative state towards higher vortex velocities as compared to the unexcited regime. Our experimental findings are largely reproduced by time-dependent Ginzburg-Landau simulations and can be explained by a competition of heating and cooling of quasiparticles escaping from vortices in conjunction with a periodic variation of their core size. Additionally, we demonstrate universality of the stimulation effect on the thermodynamic and transport properties of type II superconductors.

\section{Experimental results}

\textbf{Microwave control of the vortex dynamics}. We study the vortex dynamics under superimposed direct and microwave current drives in a coplanar waveguide (CPW) made of a Nb film with thickness $d = 50$\,nm and exhibiting a superconducting transition at $T_\mathrm{c} = 8.544$\,K. The experimental geometry is shown in Fig. \ref{fCVC}(a). The perpendicular-to-film-plane magnetic field with induction $\mathbf{B} = \mu_0\mathbf{H}$ populates the CPW with a lattice of Abrikosov vortices. The sum of applied direct and microwave currents exerts a Lorentz force on the vortices that causes their motion with velocity $v$ across the central conductor of the CPW. The associated voltage drop along the central conductor allows one to distinguish between the different resistive states of the sample.
\begin{figure*}
    \centering
    \includegraphics[width=0.8\linewidth]{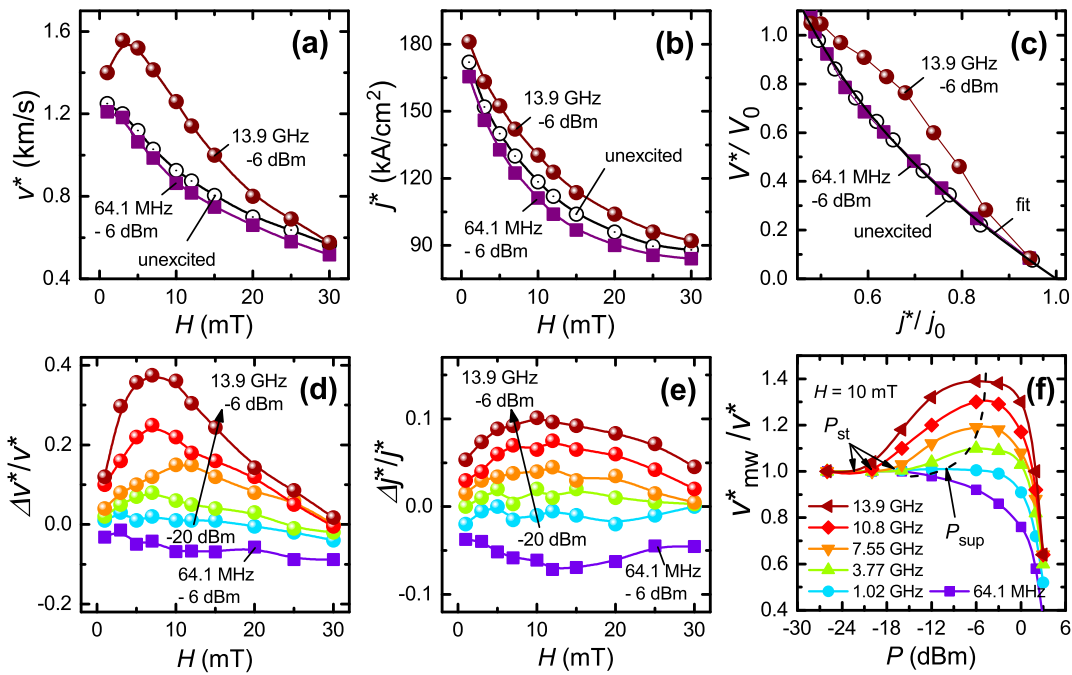}
    \caption{
    \textbf{Flux-flow instability parameters in the presence of a microwave ac stimulus}.
    (a,b) Magnetic field dependences of the velocity and the current density at the instability point at the different excitation conditions, as labeled close to the curves.
    (c) The complete set of instability points in the normalized voltage $V^\ast/V_0$ versus normalized current density $j^\ast/j_0$ representation with parameters $B_\mathrm{T}=12$\,mT,
    $V_0 = 0.155$; $0.14$; $0.17$\,V/cm and $j_0 = 184$; $176$; $192$\,kA/cm$^2$ for the unexcited state and the excitation with $64.1$\,MHz and $13.9$\,GHz, respectively.
    The solid line is a fit to the expressions (\ref{eLO}) derived within the framework of the Larkin-Ovhinnikov theory \cite{Lar75etp,Lar86inb}.
    (d,e) Relative changes of the velocity and the current density at the instability point as a function of the magnetic field value for a series of
    microwave power levels at $13.9$\,GHz in comparison with the $-6$\,dBm/$64.1$\,MHz ac excitation.
    (f) Normalized instability velocity $v^\ast_\mathrm{mw}/v^\ast$ as a function of the ac power and frequency at $10$\,mT.
    $P_\mathrm{st}$ and $P_\mathrm{sup}$ designate the microwave ac power levels above which the stability of the flux flow is stimulated and suppressed, respectively.
    In all panels $T = 0.988 T_\mathrm{c}$.
    \vspace{2mm}
    }
    \label{fVvH}
\end{figure*}

We demonstrate the control of the nonlinear resistive regime via advancing or delaying the dc-bias-induced breakdown of the non-equilibrium superconducting state. This control can be clearly seen in Fig. \ref{fCVC}(c-e) where the current-voltage ($I$-$V$) curves at $T = 0.988 T_\mathrm{c}$ are shown for a series of magnetic fields at three ac frequencies for the ac power levels $-60$\,dBm (corresponding to $1$\,nW, solid lines) and $-6$\,dBm (correspodning to $0.25$\,mW, symbols). In what follows, the excitation level $-60$\,dBm will be referred to as the unexcited state. Thus, in all $I$-$V$ curves one can recognize the nearly linear regime of flux flow followed by an upward bending at the foots of abrupt jumps to the normally conducting state. These jumps are the hallmark of the Larkin-Ovchinnikov (LO) instability \cite{Lar75etp,Lar86inb,Bez92pcs} occurring at the instability current $I^\ast$ relating to the instability voltage $V^\ast$. The definition of $I^\ast$ and $V^\ast$, as well as the respective quantities in the presence of a microwave current, $I^\ast_\mathrm{mw}$ and $V^\ast_\mathrm{mw}$, is shown in Fig. \ref{fCVC}(e).

The most striking observation in Fig. \ref{fCVC}(c-e) is that depending on the ac frequency, the microwave ac stimulus affects the onset of the flux-flow instability differently. At $64.1$\,MHz in Fig. \ref{fCVC}(c), which is exemplary for relatively low frequencies, the instability jumps occur at $I^\ast_\mathrm{mw}(H) < I^\ast(H)$. The occurrence of the instability at smaller $I$ values in the excited $I$-$V$ curves can be understood as a consequence of the replacement of $I$ by the sum of direct and microwave currents in the excited regime. By contrast, at $13.9$\,GHz, which is representative for the highest frequencies available in our experiment, the \emph{onset of the instability is shifted towards higher current values}. This shift is accompanied by the development of a pronounced upturn bending in the excited $I$-$V$ curves as compared to the unexcited ones. The expansion of the nonlinear regime in the excited $I$-$V$ curves in Fig. \ref{fCVC}(e) can be clearly seen as a deviation from the reference dashed straight line. These extended nonlinear $I$-$V$ sections at the foots of the instability jumps are the fingerprint of the stabilization effect of the microwave current on the flux flow in the investigated system. For completeness, in Fig. \ref{fCVC}(d) we present the data at $1.02$\,GHz which has been chosen as an intermediate frequency at which the excited and unexcited $I$-$V$ curves almost perfectly overlap. This coincidence can be explained by the assumption that the contribution of the microwave current into the triggering of the instability is nearly completely compensated by its stabilization effect on the flux flow.
\begin{figure*}[t!]
    \centering
    \includegraphics[width=0.65\linewidth]{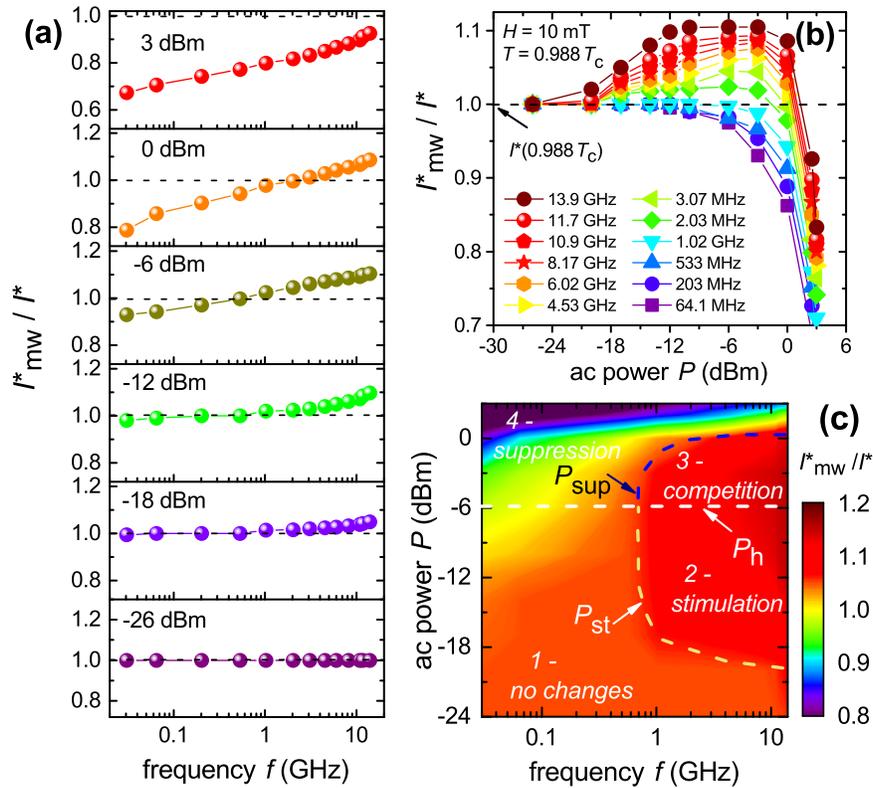}
    \caption{\textbf{Tuning the instability current by the microwave ac stimulus}.
    Normalized instability current $I^\ast_\mathrm{mw}/I^\ast$ as a function of the ac frequency for a series of microwave power levels (a)
    and as a function of the ac power for a series of ac frequencies (b), as indicated.
    (c) Contour plot $I^\ast_\mathrm{mw}(f,P)/I^\ast$. In all panels $T = 0.988T_\mathrm{c}$ and $H = 10$\,mT.
    $P_\mathrm{h}$ designates the power level above which only saturation and reduction of the instability current is observed.
    \vspace{2mm}
    }
    \label{fIvPf}
\end{figure*}

\textbf{Microwave control of instability parameters}.
To illustrate further the stabilization effect of the microwave current on the nonequilibrium state generated by moving vortices, from the last data point before the instability jump we deduce the instability velocity $v^\ast$ and the instability current density $j^\ast$ by the standard relations $v^\ast = V^\ast/(HL)$ and $j^\ast = I^\ast/(wd)$. Here, $L=1$\,mm is distance between the voltage contacts while $w=50\,\mu$m and $d=50$\,nm is the conductor width and thickness, respectively. The magnetic field dependences of the instability parameters deduced from the $I$-$V$ curves are presented in Fig. \ref{fVvH}(a) and (b). One clearly sees a decrease of $v^\ast(H)$ and $j^\ast(H)$ for each dataset, as well as $v^\ast(f)$ and $j^\ast(f)$ at a fixed $H$ value. To describe the evolution of the instability parameters in the presence of the microwave current quantitatively, we introduce new parameters $\Delta v^\ast = (v^\ast_\mathrm{mw} - v^\ast)/v^\ast$ and $\Delta j^\ast = (j^\ast_\mathrm{mw} - j^\ast)/j^\ast$ for their relative changes. Their field dependence at various microwave frequencies is shown in Fig. \ref{fVvH}(d) and (e). From the field dependences of $\Delta v^\ast(H)$ and $\Delta j^\ast(H)$ it follows that the enhancement of the critical velocity in the presence of a microwave current can reach up to $40\,\%$ while for the instability current the maximum enhancement can be about $10\,\%$. Remarkably, the enhancement of the instability parameters attains a maximum at about $10$\,mT, and the stimulation effect becomes weaker with decrease of the ac power and vanishes upon reaching an ac power level of about $-20$\,dBm. This is clearly different from the excitation of the sample with a frequency of $64.1$\,MHz at which the instability parameters are reduced by up to about $7\%$ with respect to the unexcited state.

The evolution of the critical velocity in the broad range of ac power levels and frequencies is summarized in Fig. \ref{fVvH}(f). Remarkably, one can distinguish two frequency regimes, $f \lesssim 1$\,GHz and $f \gtrsim 1$\,GHz, in which the behavior of the $v^\ast_\mathrm{mw}/v^\ast$ differs qualitatively. In the regime $f \lesssim 1$\,GHz, which we will call the low-frequency regime, the microwave current either has no effect on $v^\ast_\mathrm{mw}/v^\ast$ at lower ac power levels, or it suppresses the flux-flow stability at higher power levels, $P > P_\mathrm{sup}(f)$. By contrast, at $f \gtrsim 1$\,GHz, which will be referred to as a high-frequency regime, the $v^\ast_\mathrm{mw}/v^\ast$ ratio first remains constant, then grows in the power range $P_\mathrm{st}(f)< P < P_\mathrm{sup}(f)$, and finally decreases at yet higher microwave power levels $P > P_\mathrm{sup}(f)$. Here, $P_\mathrm{st}(f)$ and $P_\mathrm{sup}(f)$ are the frequency-dependent ac power levels above which a stimulation and a suppression effect are observed, respectively.

\begin{figure}[t!]
    \centering
    \includegraphics[width=1\linewidth]{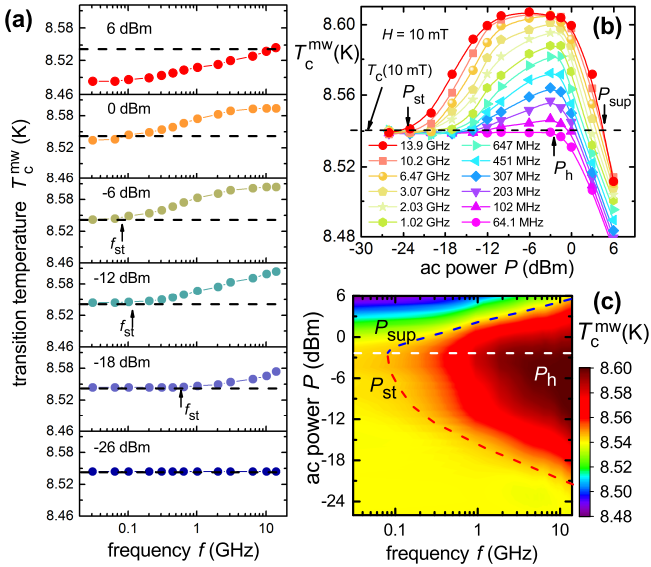}
    \caption{\textbf{Stimulation of the critical temperature by a microwave current}.
    Superconducting transition temperature $T^\mathrm{mw}_\mathrm{c}$ of the sample as a function of the ac frequency for a series of microwave power levels (a)
    and as a function of the microwave power for a series of ac frequencies (b).
    (c) Contour plot $T^\mathrm{mw}_\mathrm{c}(f,P)$. In all panels $H = 10$\,mT.
    $f_\mathrm{st}$ and $P_\mathrm{st}$ are the ac frequency and the mw power, respectively, above which $T^\mathrm{mw}_\mathrm{c}$ becomes larger than $T_\mathrm{c}$.
    $P_\mathrm{sup}$ designates the microwave power above which $T^\mathrm{mw}_\mathrm{c}$ becomes smaller than $T_\mathrm{c}$.
    }
    \label{s1}
    \vspace{0mm}
\end{figure}

\begin{figure}[t!]
    \centering
    \includegraphics[width=1\linewidth]{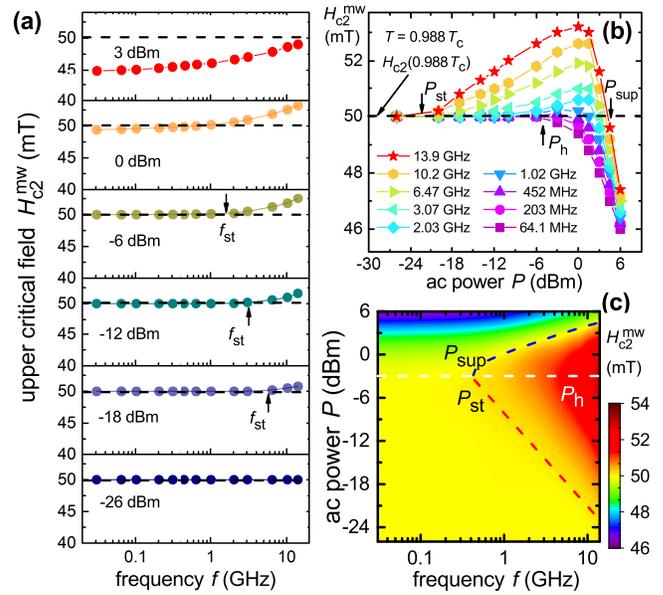}
    \caption{\textbf{Stimulation of the upper critical field by a microwave current}.
    Upper critical field $H^\mathrm{mw}_\mathrm{c2}$ of the sample as a function of the ac frequency for a series of microwave power levels (a)
    and as a function of the microwave power for a series of ac frequencies (b).
    (c) Contour plot $H^\mathrm{mw}_\mathrm{c2}(f,P)$. In all panels $T = 0.988 T_\mathrm{c}$.
    $f_\mathrm{st}$ and $P_\mathrm{st}$ are the ac frequency and the microwave power, respectively, above which $H^\mathrm{mw}_\mathrm{c2}$ becomes larger than $H_\mathrm{c2}$.
    $P_\mathrm{sup}$ designates the microwave power above which $H^\mathrm{mw}_\mathrm{c2}$ becomes smaller than $H_\mathrm{c2}$.
    }
    \label{s2}
    \vspace{2mm}
\end{figure}

A similar stimulation effect for the maximum current $I^\ast_\mathrm{mw}$, up to which the low-resistive state is maintained in the presence of a microwave ac stimulus, is demonstrated in Fig. \ref{fIvPf}. In Fig. \ref{fIvPf}(a) the frequency dependence of the normalized current $I^\ast_\mathrm{mw}/I^\ast$ exhibits a systematic evolution from a weak frequency dependence at low ac power levels to a notably growing tendency at higher power levels. While the dashed lines in each panel in Fig. \ref{fIvPf}(a) depict the reference level $I^\ast_\mathrm{mw}/I^\ast=1$ in the absence of a microwave excitation, an enhancement of the instability current is clearly seen at higher frequencies and power levels exceeding about $-12$\,dBm. At the same time, $I^\ast_\mathrm{mw}/I^\ast$ begins to decrease with a further increase of the microwave power. The dependence $I^\ast_\mathrm{mw}(P,f)/I^\ast$ in Fig. \ref{fIvPf}(b) and (c) is qualitatively very similar to that for $v^\ast_\mathrm{mw}(P,f)/v^\ast$ in Fig. \ref{fVvH}(f). Furthermore, both dependences are very similar to those for the superconducting transition temperature $T^\mathrm{mw}_\mathrm{c}(P,f)/T_\mathrm{c}$ and the upper critical field $H^\mathrm{mw}_\mathrm{c2}(P,f)/H_\mathrm{c2}$ in Figs. \ref{s1} and \ref{s2}. Interestingly, the maximal increase of the transition temperature in the presence of an ac stimulus is only of the order of $1\%$, while that for $H_\mathrm{c2}$ reaches about $6\%$. At the same time, the relative enhancement of the instability velocity and the instability current is significantly larger and reaches up to $40\%$ and $10\%$, respectively.

The effect of a microwave ac stimulus on the critical transport and thermodynamic parameters can be summarized in the $f$-$P$ phase diagram of the instability current in Fig. \ref{fIvPf}(c), where four different regimes can be identified. Namely, at low ac frequencies and power levels there is virtually no effect of the microwave stimulus, as expected (region 1 in Fig. \ref{fIvPf}(c)). With increasing ac power, at $\gtrsim 1$\,GHz an enhancement of the parameters is observed (region 2 in Fig. \ref{fIvPf}(c)). At a further increase of the ac power up to $P_\mathrm{h}$ the parameters at higher frequencies stop to increase even further, while at low frequencies the parameters begin to decrease (region 3 in Fig. \ref{fIvPf}(c)). Finally, at high ac power levels a suppression of the superconducting critical parameters is observed regardless of the ac frequency (region 4 in Fig. \ref{fIvPf}(c)).

\section*{Simulation results}
\begin{figure*}
    \centering
    \includegraphics[width=0.7\linewidth]{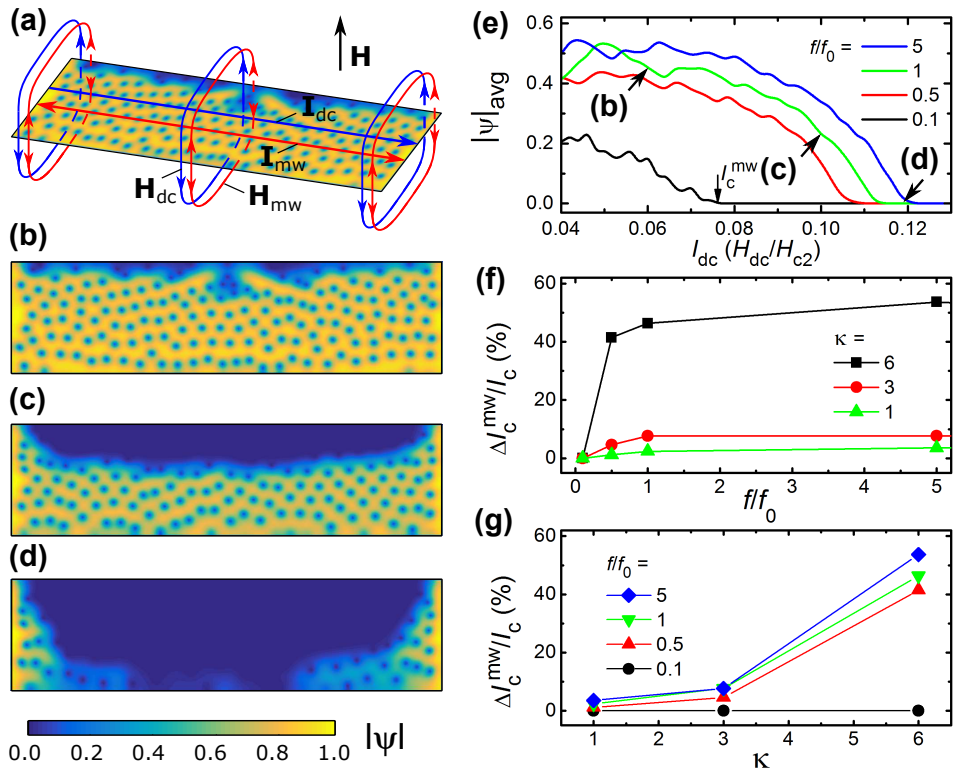}
    \caption{
    \textbf{Time-dependent Ginzburg-Landau simulations}.
    (a) Sketch of the simulated system: The external field $\mathbf{H}$ is uniform in all cells while $\mathbf{H}_\mathrm{dc}$ and $\mathbf{H}_\mathrm{mw}$ only exist along the conductor boundaries being generated by the dc and microwave currents, respectively, as depicted by the arrows.
    (b)-(d) Contour plots of the order parameter $|\psi|$ at different dc current values, as indicated in panel (e).
    The data in (f) are calculated at $T = 0.47T_\mathrm{c}$, $H = 0.2H_\mathrm{c2}$, and $I_\mathrm{mw} $ corresponding to $0.2H_\mathrm{c2}$.
    Relative change of the critical current in the presence of a microwave stimulus, $\Delta I^\mathrm{mw}_\mathrm{c} \equiv I^\mathrm{mw}_\mathrm{c} - I_\mathrm{c}$, normalized to its magnitude in the absence of a microwave excitation $I_\mathrm{c}$, as a function of the dimensionless ac frequency (f) and for a series of the Ginzburg-Landau parameter $\kappa$ (g).
        \vspace{2mm}
    }
    \label{fTDGL1}
\end{figure*}

Since analytical theories of microwave-stimulated mixed state and the flux-flow instability in the presence of a high-frequency ac current are unavailable so far, further insights into the evolution of the dynamic state generated by vortex motion can be gained on the basis of computer simulations relying upon the solution of the time-dependent Ginzburg-Landau (TDGL) equation. While the microscopic derivation of the TDGL was originally done for a gapless superconductor with paramagnetic impurities \cite{Gor68etp}, it is also widely used for studying various aspects of current-driven vortex matter in superconductors with gap and non-magnetic impurities \cite{Ara02rmp,Kwo16rpp}, including the vortex dynamics at high vortex velocities and at GHz ac frequencies \cite{Her08prb,Emb17nac,Ori19arx}. In the TDGL, not only the vortex-vortex interaction is taken into account, but also vortex-core structures and interaction with pinning centers can be described. At the same time, far from $T_\mathrm{c}$, the TDGL equation does not reproduces the physics in the vortex core quantitatively, but it still describes the spatiotemporal evolution of vortex matter qualitatively \cite{Kwo16rpp}. For the simulations we adopt the link variable method \cite{Lar15nsr,Gro96jcp} for 2D and 3D systems of adjustable size and shape and use a solver that approximates numerically the solution of the TDGL equation. The gauge for the TDGL solver is such that the scalar potential is zero, so no electric currents can be introduced directly in the simulations and the effect of the currents is introduced by the stray field they would generate. The simulations take place in a two-dimensional $360\times100$ cell superconducting strip, with a constant and uniform external field $H=0.2H_\mathrm{c2}$ applied perpendicular to the plane of the strip. The magnetic field $\mathbf{H}_\mathrm{dc}$ of opposite directions at each boundary is applied to simulate the field generated by the dc bias current. The magnitude of $H_\mathrm{dc}$ is varied in steps of $0.005H_\mathrm{c2}$ every $10^4$ steps of the simulation, giving enough time to reach a stationary state before increasing $H_\mathrm{dc}$ further. Finally, an additional field $\mathbf{H}_\mathrm{mw}$ is applied with opposite and alternating directions at each boundary, simulating the high-frequency ac current of constant amplitude and frequency $f$, that affects the dynamic state of the strip. The geometry of the applied fields is illustrated in Fig. \ref{fTDGL1}(a). More details on the simulation procedure are given in the Methods section.

\begin{figure}[t!]
    \centering
    \includegraphics[width=1\linewidth]{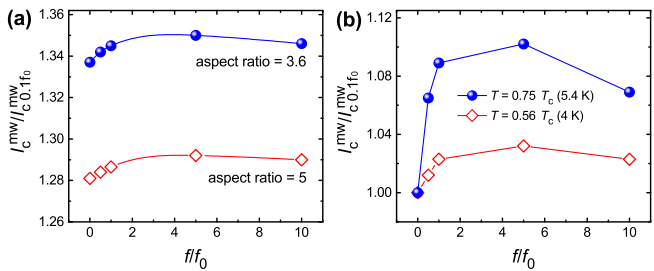}
    \caption{
    \textbf{Dependence of the critical current on the microstrip aspect ratio and temperature}.
    (a) The instability current $I^\mathrm{mw}_\mathrm{c}$ normalized to its value in the presence of an ac current with frequency $0.1f_0$, $I^\mathrm{mw}_\mathrm{c\,0.1f_0}$, as a function of the ac frequency for two different ratios of the microstrip length and width (a) and two temperature values (b).
    }
    \label{fTDGL2}
\end{figure}

The simulations are done for Nb with the Ginzburg-Landau parameter $\kappa = 6$ deduced from the experiment and the critical temperature $T_\mathrm{c}=8.5$\,K. With these values, simulations were performed at $T=4$\,K with ac frequencies $f=0.1f_0$, $0.5f_0$, $f_0$, $5f_0$ and $10f_0$ and with an ac field amplitude of $0.2H_\mathrm{c2}$. Here, $f_0=1/t_0$, where $t_0=\pi\hbar/96k_BT_\mathrm{c}$ is the characteristic time of the relaxation of the order parameter \cite{Kat93prb}. The low temperature was chosen for the only reason of a better contrast in the spatial dependence of the order parameter in the simulations of the vortex patterns. After setting the temperature, the external magnetic field and the main parameters defining the material, the order parameter is initialized in the superconducting state $|\psi|=1$ and its evolution is simulated as a function of the dc current in the presence of a high-frequency ac current stimulus, Fig. \ref{fTDGL1}(b). Soon after start of the simulations the Nb strip is quickly filled with vortices, due to the applied external field. As soon as the dc current begins to increase, there appears a net flux of vortices in the direction of the $y$ axis, from top to bottom. A further increase of the dc current causes a shift of vortices that turns the upper part of the strip into the normal state, and eventually breaks the superconducting channel. The different stages of this process can be seen in Fig. \ref{fTDGL1}(c) illustrating the vortex patterns and the evolution of the superconducting channel with increasing dc current value. In the simulations, the critical current $I_\mathrm{c}$ is defined as a current at which the superconducting channel breaks down. The relative variation of this current is measured as a function of the normalized drive frequency $f$, exhibiting a $60\%$ increase as the frequency is varied from $0.1f_0$ to $10f_0$. The increase of $I_\mathrm{c}(f)$ is nonlinear, with a faster increase between $0.1f_0$ and $f_0$, see Fig. \ref{fTDGL1}(f). We have also checked that the ac-stimulated increase of $I_\mathrm{c}$ varies with $\kappa$, as can be seen in Fig. \ref{fTDGL1}(g). Thus, for $\kappa=1$ the variation is smaller than about $3.5\,\%$, while for $\kappa=6$ it can reach $60\,\%$.
The dependences of the critical current on the microstrip aspect ratio and temperature are illustrated in Fig. \ref{fTDGL2}.

\section*{Discussion}
The reported experimental findings relate to a fundamental problem of the interaction of a transport current containing dc and high-frequency ac components with a superconductor in the vortex state. As is well known, the vortex state is characterized by a spatially modulated superconducting order parameter which vanishes in the vortex cores and attains a maximal value between them. Accordingly, a type II superconductor can be regarded as a continuum medium consisting of bunches of quasiparticles in the vortex cores surrounded by a bath of the superconducting condensate formed by the superfluid of Cooper pairs  \cite{Bra95rpp}. To the authors' best knowledge, so far there is no available theory addressing the complex interaction of the superimposed dc and microwave currents with the quasiparticles and the condensate at the same time. Yet, some ingredients of this interaction, which were studied separately so far, should be mentioned prior to discussing the experimental findings.

The effect of a dc current on the superconducting condensate is well known  \cite{Ant03prl}. With an increase of a dc current, the absolute value of the order parameter is decreasing and the peak in the density of states at the edge of the superconducting gap is smeared. This is because of gaining a finite momentum by the Cooper pairs that form a coherent excited state that plays a central role in the explanation of the gauge invariance of the Meissner effect  \cite{And58prv}. The equivalence of depairing due to an electric current and due to a magnetic field is also well known, both theoretically  \cite{Mak65prv} and experimentally  \cite{Ant03prl}, and we have in fact used this equivalence in the TDGL simulations.

A general theory of depairing by a microwave field was formulated quite recently\, \cite{Sem16prl}. It was shown that the ground state of a superconductor is altered qualitatively in analogy to the depairing due to a dc current. However, in contrast to dc depairing, the density of states acquires steps at multiples of the microwave photon energy and shows an exponential-like tail in the subgap regime  \cite{Sem16prl}. Additionally, depending on temperature, one can consider two regimes in which the response of a superconductor is dominated either by the response of the superfluid (at low temperatures, $T/T_\mathrm{c}\ll 1$) or by the quasiparticles (close to the critical temperature, $(T-T_\mathrm{c})/T_\mathrm{c}\ll 1$, as in our experiment). It is also known that at $T\lesssim T_\mathrm{c}$, microwave radiation can be absorbed by quasiparticles, leading to a nonequilibrium distribution over the energies \cite{Eli70etp}.

In general, the response of the condensate to an external microwave field becomes apparent via a change of the kinetic impedance (imaginary part of the complex resistivity) while the quasiparticles give rise to the microwave loss (real part of the complex resistivity). In the presence of an external magnetic field inducing vortices in the superconductor, the vortex-induced resistive loss dominates the response of the superconductor. We note that the ac frequencies in our experiment are much smaller than the superconducting gap frequency $f\ll f_\mathrm{gap}(0.998T_\mathrm{c}) \simeq 75$\,GHz and the experiment is done in the vicinity of $T_\mathrm{c}$ where the superconducting gap is small and the GL, LO, and Eliashberg theories are justified \cite{Eli70etp,Moo81inb,Lar76etp,Lar86inb,Gor68etp,Kwo16rpp}.

It is important to stress that the known effects of stimulation of superconductivity by a microwave stimulus in the absence of an external magnetic field include an enhancement of the Ginzburg-Landau depairing current (in narrow channels) implying a transition to a resistive state due to the formation of phase-slip centers  \cite{Zol13ltp} or the Aslamazov-Lempitskii maximum current (in wide films) at which the vortex structure induced by the self-field evolves into the first phase-slip line  \cite{Asl82etp}. By contrast, in the presence of an external magnetic field when Abrikosov vortices move under the action of the transport current, there is an additional phenomenon leading to an abrupt quenching of the superconductor to the normally conducting state \emph{earlier than the Ginzburg-Landau or Aslamazov-Lempitskii critical current is reached}. Within the LO theoretical framework \cite{Lar75etp,Lar86inb}, this quenching is because of the flux-flow instability caused by the nonlinear dependence of the film conductivity on the electric field. However, the LO theory was developed in the dirty limit near $T_\mathrm{c}$ and for weak magnetic fields when the heating of the superconductor can be neglected. To account for a finite density of vortices, the LO theory was extended by Bezuglyj and Shklovskij in Ref. \cite{Bez92pcs}. In that work \cite{Bez92pcs}, the complete set of instability points in $I$-$V$ curves for a series magnetic field values is described by the system of equations
\begin{equation}
\label{eLO}
    \begin{array}{lll}
    \displaystyle\frac{E^\ast}{E_0}=\frac{(1-t)(3t+1)}{2\sqrt{2}t^{3/4}(3t-1)^{1/2}},
    \\[4mm]
    \displaystyle\frac{j^\ast}{j_0} = \frac{2\sqrt{2}t^{3/4}(3t-1)^{1/2}}{3t+1},
    \end{array}
\end{equation}
where $t=[1+b+(b^2 +8b+4)^{1/2}]/3(1+2b)$ and $b = B / B_\mathrm{T}$ is the dimensionless magnetic field with the parameter
\begin{equation}
\label{eBt}
    B_\mathrm{T} = 0.374 k_\mathrm{B}^{-1}c e_0 R_{\square}h\tau_\varepsilon.
\end{equation}
Here, $k_\mathrm{B}$ is the Boltzmann constant, $c$ the speed of light in vacuum, $R_{\square} = (\sigma_n d)^{-1}$ the film sheet resistance, $e_0$ the electron charge, $h$ the heat removal coefficient, and $\tau_\varepsilon$ the quasiparticle energy relaxation time. In the system of equations (\ref{eLO}), the parameters $E_0$ and $j_0$ are defined as
\begin{equation}
\label{eLOparam}
    \begin{array}{lll}
    \displaystyle E_0 = 1.02(B_\mathrm{T}/c)(D/\tau_\varepsilon)^{1/2}(1-T_\mathrm{B}/T_\mathrm{c})^{1/4},
    \\[4mm]
    \displaystyle j_0 = 2.62 (\sigma_\mathrm{n}/e_0)(D\tau_\varepsilon)^{-1/2}k_\mathrm{B} T_\mathrm{c}(1-T_\mathrm{B}/T_\mathrm{c})^{3/4}.
    \end{array}
\end{equation}

The curve calculated by Eqs. (\ref{eLO}) and (\ref{eLOparam}) is shown in Fig. \ref{fVvH}(c) which also contains the experimentally measured instability points in the normalized voltage $V^\ast/V_0\equiv E^\ast/E_0$ versus normalized current density $j^\ast/j_0$ representation. The normalization parameters used for fitting the experimental data to the theoretical curve are $B_\mathrm{T}=12$\,mT with $V_0 = 0.155$; $0.14$; $0.17$\,V/cm and $j_0 = 184$; $176$; $192$\,kA/cm$^2$ for the unexcited state and the excitation with $64.1$\,MHz and $13.9$\,GHz, respectively. From the figure it follows that the instability points $E^\ast(j^\ast/j_0)/E_0$ in the absence of ac current and in the presence of an ac current with $f=64.1$\,MHz nicely fit to the expressions (\ref{eLO}). By contrast, in the presence of an ac current with $f=13.9$\,GHz a noticeable deviation from the expressions (\ref{eLO}) is observed. While it was recently revealed that possible inhomogeneities in the distribution and strength of pinning sites do not alter the dependence given by Eq. (\ref{eLO}) qualitatively \cite{Bez19prb}, but rather require a renormalization of the parameters $V_0$ and $E_0$, our findings suggest that the superimposed microwave ac current at high enough power levels becomes a crucial ingredient which qualitatively modifies the physical picture and causes a deviation from the flux-flow instability theory developed for the sole case of a dc bias current. Namely, the microscopic scenario of the flux-flow instability under a dc current drive implies a decrease in the number of quasiparticles in the vortex cores under the action of an electric field. In return, the decrease in the number of quasiparticles leads to a shrinkage of the vortex cores and a decrease in the vortex viscosity with increasing vortex velocity. As a consequence, the viscous force has a maximum as a function of the vortex velocity, and as soon as the Lorentz force exceeds this maximum, the viscous flow of the vortices becomes unstable.

From the specific power at the instability point, $P_0 = j_0 E_0 =(h/d)(T_\mathrm{c} - T)$ \cite{Bez92pcs,Lef99pcs,Per05prb,Bez19prb}, following from Eqs. (\ref{eLOparam}) with $\sigma_n = 1/(R_\square d)$, one can deduce the heat removal coefficient $h\approx 1.2$\,WK$^{-1}$cm$^{-2}$. Substitution of $h$ and $B_\mathrm{T}=12$\,mT into Eq. (\ref{eBt}) yields the energy relaxation time $\tau_{\varepsilon} \approx 0.23$\,ns. For the relaxation time associated with the electron-phonon scattering in the LO model, we find this estimate to be in reasonable agreement with the $\tau_{\varepsilon}$ estimates of $0.15$\,ns \cite{Kap76prb}, $0.2$-$0.4$\,ns \cite{Per05prb}, and $0.3$-$0.7$\,ns \cite{Leo11prb} for Nb thin films. We note that if one uses the order parameter relaxation time ($1.8\cdot10^{-11}$\,s for Nb \cite{Kap76prb,Moo81inb}) for the estimate of the minimal frequency $f_\mathrm{min} = 1.73/(2 \pi \tau_\varepsilon)$ \cite{Moo81inb}, above which an enhancement of superconductivity via the Eliashberg mechanism is possible, than no microwave stimulation effect is expected at $f < f_\mathrm{min}\thickapprox15$\,GHz \cite{Moo81inb}. However, for gapped superconductors, such as Nb, it was argued that the gap change is much slower, with a \emph{relaxation rate dominated by the electron-phonon scattering} \cite{Sch68pkm,Tin04boo}. Accordingly, if one takes $\tau_{\varepsilon} = 0.23$\,ns for the estimate of the threshold frequency in Nb, one obtains $f_\mathrm{min} \approx 1.2$\,GHz. While the microscopic description of the studied system should clearly go beyond the Eliashberg theory \cite{Eli70etp,Moo81inb}, we underline that the enhancement of the critical parameters in our experiment is observed at $f \gtrsim 1$\,GHz. In this way, the electron-phonon interaction time appears to be an important ingredient governing microwave-stimulated superconductivity in the vortex state in Nb films.

To elucidate the extension of the low-resistive flux-flow regime in the presence of (dc+ac) current drives, we suggest the following qualitative explanation. Namely, when an ac current is added to the dc current, at rather high microwave power levels the vortices start oscillating near their equilibrium positions, which in return are displaced due to vortex motion under the action of the dc driving current. In other words, the dc current leads to a \emph{translational motion} of vortices while the ac stimulus induces a \emph{breathing mode} in addition to their translation. This breathing mode appears due to the variation of the vortex core sizes in time due to the periodically-modulated quasiparticle escape from the cores. Accordingly, the strong oscillations of vortices are expected to lead to the formation of ``clouds'' of quasiparticles around the vortex cores, whose relaxation should now take place in a \emph{larger volume} as compared to the non-excited case. At the same time, the dissipation is decreasing due to the combined effects of the shrinkage of the vortex cores and the redistribution of quasiparticles away from the gap edge via the Eliashberg mechanism. While a detailed theory of microwave-stimulated superconductivity in the vortex state at high vortex velocities is yet to be elaborated, we believe that the model of ``moving breathing hot spots'' can be applied for a qualitative explanation of the observed effects.

To summarize, we have investigated the effect of an ac current with frequencies in the MHz and the lower GHz range on the resistive state of superconducting Nb films subjected to a dc bias current in perpendicular magnetic fields. While without ac excitation and in the presence of an ac excitation at frequencies in the MHz range the quenches of the superconductor to a highly-resistive state are nicely described by the expressions derived within the framework of the Larkin-Ovchinnikov theory of flux-flow instability, the addition of a GHz-frequency ac current at moderately high power levels leads to a notable deviation from the Larkin-Ovchinnikov theory and to an extension of the low-resistive dynamical state to larger current values. This key experimental observation is supported by simulations based on the time-dependent Ginzburg-Landau equation and can be explained qualitatively by a model of ``breathing mobile hot spots'' implying a competition of heating and cooling of quasiparticles along the trajectories of moving fluxons whose core sizes vary in time. While a complete theory of microwave-stimulated superconductivity in the vortex state remains unavailable so far, our results are relevant for superconducting microwave circuits exploited in quantum computing and remote sensing, as well as they furthermore might have implications for diverse physical problems allowing mapping of their solutions to scalar condensates.
\vspace{-0.2cm}

\section*{Methods}
\vspace{-0.2cm}
\subsection{Film growth and characterization.}
The CPWs were fabricated by photolithography and Ar etching from epitaxial (110) Nb films on a-cut sapphire substrates. The films were grown by dc magnetron sputtering in a setup with a base pressure in the $10^{-8}$\,mbar range. The films were sputtered at the substrate temperature $T =800^\circ$C. The Ar pressure was $5\times10^{-3}$\,mbar and the film growth rate was $1$\,nm/s. The (110) orientation of the films was inferred from X-ray diffraction measurements \cite{Dob12tsf}. The epitaxy of the films was confirmed by reflection high-energy electron diffraction. The as-grown films have a smooth surface with an rms surface roughness of less than $0.5$\,nm, as inferred from atomic force microscopy inspection in the range $1\,\mu$m$\times1\,\mu$m. The films are characterized by a superconducting transition temperature $T_\mathrm{c}$ in zero field of $8.544$\,K, as determined by the $90\,\%$ resistance criterion. Their normal-state resistivity just above the superconducting transition amounts to $2.48\,\mu\Omega$cm and the upper critical field at zero temperature is estimated as $H_\mathrm{c2}(0) = H_\mathrm{c2}(T)/[1 - (T/T_\mathrm{c})^2] \approx 1.2$\,T corresponding to a superconducting coherence length $\xi(0) = (\Phi_0 /2\pi H_\mathrm{c2})^{1/2} \approx 17$\,nm. Here, $\Phi_0$ is the magnetic flux quantum. The magnetic field penetration depth $\lambda(0)$ at zero temperature in the films can be estimated as 100\,nm \cite{Gub05prb} yielding the Ginzburg-Landau parameter $\kappa\approx 6$.

\subsection{Microwave and dc electrical measurements.}
Combined broadband microwave and dc electrical measurements were done in a $^4$He cryostat with magnetic field $\mathbf{H}$ directed perpendicular to the film surface. A custom-made cryogenic sample probe with coaxial cables was employed. The microwave and dc currents were superimposed and uncoupled by using two bias-tees mounted at the VNA ports. The dc voltage measurements were performed employing a Keithley Sourcemeter 2635B and an Agilent 34420A nanovoltmeter. The microwave signal, with a frequency between $30$\,MHz and $14$\,GHz, was generated and analyzed by a Keysight-Agilent E5071C vector network analyzer (VNA). In the transmission line, the microwave power level $-6$\,dBm ($0.25$\,mW) corresponds to the nominal microwave current $\simeq2.2$\,mA. The sample under study is a $50\,$Ohm-matched Nb coplanar waveguide (CPW), with the width of the central conductor $w = 50\,\mu$m, the center-to-ground distance $a = 20\,\mu$m and the length of the active area $L = 1$\,mm. The coaxial cables were connected to the CPW via SMPB connectors spring-loaded to the gold-plated contact pads sputtered on top of the Nb film using a shadow mask. With the estimate for the superconducting gap frequency $f_\mathrm{gap} \approx 700$\,GHz \cite{Leh94apl,Pro98prb} and the weak-coupling temperature dependence of the superconducting gap, $\Delta(T) = \Delta(0)(1- T/T_\mathrm{c})^{1/2}$, our experiments are in the subgap excitation regime with microwave frequencies $f \ll f_\mathrm{gap}(0.998T_\mathrm{c}) \simeq 75$\,GHz.

\subsection{Time-dependent Ginzburg-Landau simulations.}
In the simulations, we solve numerically the Time Dependent Ginzburg-Landau (TDGL) equations \cite{Kwo16rpp,Ara02rmp} using the link variable method. To accomplish this, 2D and 3D TDGL solvers were developed following Ref. \cite{Bus00inb}. When testing the first simulations and comparing the results, both solvers gave very similar results on vortex motion and critical currents. Because of this, the 2D solver was used since it requires considerably less computational effort.

The simulations take place in a 360x100 cells domain of superconducting material. The size of the cells is scaled by the coherence length $\xi_0=\xi(0)$, which was experimentally estimated to be around $\xi_0=17$ nm. The lateral size of each cell is $0.5\xi_0$, so the simulated strip is about $3\times0.8\,\mu$m$^2$. In order to simulate a strip such as the one used in the experiments, we could either make the cells represent a much bigger area or use a domain with $\sim10^5$ cells on each side. The former would result in a simulation unable to resolve individual vortices, and the latter would take huge amounts of time and computational resources for a single simulation. Our solution to this was choosing a size such that we can both have a big number of vortices (more than 100 vortices fit easily in each simulation, as can be seen in Fig. \ref{fTDGL1}) and resolve their individual shape and motion, while being able to perform several simulations in a reasonable time. Therefore, these simulations must be interpreted in qualitative terms to understand the effects of different dc and ac currents on superconductivity, and not as a quantitative prediction.

The simulation procedure is as following: An external dc magnetic field, uniform everywhere, is applied and populates the sample with vortices. An ac magnetic field (generated by the ac current) of different sign is applied at each boundary, of constant amplitude and frequency throughout each simulation. Finally, a dc magnetic field, of different signs at each boundary, is applied. Its magnitude is changed in steps of $0.005 H_\mathrm{c2}$ every $10^4$ steps, with the time step defined as
\begin{equation}
dt=0.9\times\frac{1}{2\kappa^2\left(\displaystyle\frac{1}{{a_x}^2}+\displaystyle\frac{1}{{a_y}^2}+\displaystyle\frac{1}{{a_z}^2}\right)},
\end{equation}
$a_{x,y,z}=0.5\xi_0$ being the lateral size of each cell, and time in units of the relaxation time for the order parameter, $t_0=\pi\hbar/96k_BT_\mathrm{c}$. For the smallest time step used (using $\kappa=6$, thus $dt=0.00104 t_0$) this equals to $10^4\times 0.00104=10.4 t_0$, giving enough time to reach a stationary state before this dc stray field was increased once more. This is similar to changing the applied dc current in the current-voltage curve. More details on the simulation procedure are provided in Ref. \cite{Lar20ltp}.

O.V.D. acknowledges the German Research Foundation (DFG) for support through Grant No 374052683 (DO1511/3-1).
C.G.-R., A.L., and F.G.A. acknowledge support by Spanish Ministerio de Ciencia (MAT2015-66000-P, RTI2018-095303-B-C55, EUIN2017-87474, MDM-2014-0377) and Consejeria de Educacion e Investigacion de la Comunidad de Madrid (NANOMAGCOST-CM Ref. P2018/NMT-4321) Grants.
V.M.B., V.A.S., A.I.B., and R.V.V. acknowledge support from the European Commission in the framework of the program Marie
Sklodowska-Curie Actions --- Research and Innovation Staff Exchange (MSCA-RISE) under Grant Agreement No. 644348 (MagIC).
Research leading to these results was conducted within the framework of the COST Action CA16218 (NANOCOHYBRI) of the European Cooperation in Science and Technology.
Support through the Frankfurt Center of Electron Microscopy (FCEM) is gratefully acknowledged.


\begin{thebibliography}{80}%
\makeatletter
\providecommand \@ifxundefined [1]{%
 \@ifx{#1\undefined}
}%
\providecommand \@ifnum [1]{%
 \ifnum #1\expandafter \@firstoftwo
 \else \expandafter \@secondoftwo
 \fi
}%
\providecommand \@ifx [1]{%
 \ifx #1\expandafter \@firstoftwo
 \else \expandafter \@secondoftwo
 \fi
}%
\providecommand \natexlab [1]{#1}%
\providecommand \enquote  [1]{``#1''}%
\providecommand \bibnamefont  [1]{#1}%
\providecommand \bibfnamefont [1]{#1}%
\providecommand \citenamefont [1]{#1}%
\providecommand \href@noop [0]{\@secondoftwo}%
\providecommand \href [0]{\begingroup \@sanitize@url \@href}%
\providecommand \@href[1]{\@@startlink{#1}\@@href}%
\providecommand \@@href[1]{\endgroup#1\@@endlink}%
\providecommand \@sanitize@url [0]{\catcode `\\12\catcode `\$12\catcode
  `\&12\catcode `\#12\catcode `\^12\catcode `\_12\catcode `\%12\relax}%
\providecommand \@@startlink[1]{}%
\providecommand \@@endlink[0]{}%
\providecommand \url  [0]{\begingroup\@sanitize@url \@url }%
\providecommand \@url [1]{\endgroup\@href {#1}{\urlprefix }}%
\providecommand \urlprefix  [0]{URL }%
\providecommand \Eprint [0]{\href }%
\providecommand \doibase [0]{http://dx.doi.org/}%
\providecommand \selectlanguage [0]{\@gobble}%
\providecommand \bibinfo  [0]{\@secondoftwo}%
\providecommand \bibfield  [0]{\@secondoftwo}%
\providecommand \translation [1]{[#1]}%
\providecommand \BibitemOpen [0]{}%
\providecommand \bibitemStop [0]{}%
\providecommand \bibitemNoStop [0]{.\EOS\space}%
\providecommand \EOS [0]{\spacefactor3000\relax}%
\providecommand \BibitemShut  [1]{\csname bibitem#1\endcsname}%
\let\auto@bib@innerbib\@empty
\bibitem [{\citenamefont {Barone}\ and\ \citenamefont
  {Patterno}(1982)}]{Bar82boo}%
  \BibitemOpen
  \bibfield  {author} {\bibinfo {author} {\bibfnamefont {A.}~\bibnamefont
  {Barone}}\ and\ \bibinfo {author} {\bibfnamefont {G.}~\bibnamefont
  {Patterno}},\ }\href@noop {} {\emph {\bibinfo {title} {Physics and
  Applications of the Josephson Effect}}}\ (\bibinfo  {publisher} {John Wiley
  \& Sons, New York},\ \bibinfo {year} {1982})\BibitemShut {NoStop}%
\bibitem [{\citenamefont {Welp}\ \emph {et~al.}(2013)\citenamefont {Welp},
  \citenamefont {Kadowaki},\ and\ \citenamefont {Kleiner}}]{Wel13nph}%
  \BibitemOpen
  \bibfield  {author} {\bibinfo {author} {\bibfnamefont {U.}~\bibnamefont
  {Welp}}, \bibinfo {author} {\bibfnamefont {K.}~\bibnamefont {Kadowaki}}, \
  and\ \bibinfo {author} {\bibfnamefont {R.}~\bibnamefont {Kleiner}},\ }\href
  {http://dx.doi.org/10.1038/nphoton.2013.216} {\bibfield  {journal} {\bibinfo
  {journal} {Nat. Photon.}\ }\textbf {\bibinfo {volume} {7}},\ \bibinfo {pages}
  {702} (\bibinfo {year} {2013})}\BibitemShut {NoStop}%
\bibitem [{\citenamefont {Dobrovolskiy}\ \emph {et~al.}(2018)\citenamefont
  {Dobrovolskiy}, \citenamefont {Bevz}, \citenamefont {Mikhailov},
  \citenamefont {Yuzephovich}, \citenamefont {Shklovskij}, \citenamefont
  {Vovk}, \citenamefont {Tsindlekht}, \citenamefont {Sachser},\ and\
  \citenamefont {Huth}}]{Dob18nac}%
  \BibitemOpen
  \bibfield  {author} {\bibinfo {author} {\bibfnamefont {O.~V.}\ \bibnamefont
  {Dobrovolskiy}}, \bibinfo {author} {\bibfnamefont {V.~M.}\ \bibnamefont
  {Bevz}}, \bibinfo {author} {\bibfnamefont {M.~Y.}\ \bibnamefont {Mikhailov}},
  \bibinfo {author} {\bibfnamefont {O.~I.}\ \bibnamefont {Yuzephovich}},
  \bibinfo {author} {\bibfnamefont {V.~A.}\ \bibnamefont {Shklovskij}},
  \bibinfo {author} {\bibfnamefont {R.~V.}\ \bibnamefont {Vovk}}, \bibinfo
  {author} {\bibfnamefont {M.~I.}\ \bibnamefont {Tsindlekht}}, \bibinfo
  {author} {\bibfnamefont {R.}~\bibnamefont {Sachser}}, \ and\ \bibinfo
  {author} {\bibfnamefont {M.}~\bibnamefont {Huth}},\ }\href {\doibase
  10.1038/s41467-018-07256-0} {\bibfield  {journal} {\bibinfo  {journal} {Nat.
  Commun.}\ }\textbf {\bibinfo {volume} {9}},\ \bibinfo {pages} {4927}
  (\bibinfo {year} {2018})}\BibitemShut {NoStop}%
\bibitem [{\citenamefont {Marsili}\ \emph {et~al.}(2013)\citenamefont
  {Marsili}, \citenamefont {Verma}, \citenamefont {Stern}, \citenamefont
  {Harrington}, \citenamefont {Lita}, \citenamefont {Gerrits}, \citenamefont
  {Vayshenker}, \citenamefont {Baek}, \citenamefont {Shaw}, \citenamefont
  {Mirin},\ and\ \citenamefont {Nam}}]{Mar13nph}%
  \BibitemOpen
  \bibfield  {author} {\bibinfo {author} {\bibfnamefont {F.}~\bibnamefont
  {Marsili}}, \bibinfo {author} {\bibfnamefont {V.~B.}\ \bibnamefont {Verma}},
  \bibinfo {author} {\bibfnamefont {J.~A.}\ \bibnamefont {Stern}}, \bibinfo
  {author} {\bibfnamefont {S.}~\bibnamefont {Harrington}}, \bibinfo {author}
  {\bibfnamefont {A.~E.}\ \bibnamefont {Lita}}, \bibinfo {author}
  {\bibfnamefont {T.}~\bibnamefont {Gerrits}}, \bibinfo {author} {\bibfnamefont
  {I.}~\bibnamefont {Vayshenker}}, \bibinfo {author} {\bibfnamefont
  {B.}~\bibnamefont {Baek}}, \bibinfo {author} {\bibfnamefont {M.~D.}\
  \bibnamefont {Shaw}}, \bibinfo {author} {\bibfnamefont {R.~P.}\ \bibnamefont
  {Mirin}}, \ and\ \bibinfo {author} {\bibfnamefont {S.~W.}\ \bibnamefont
  {Nam}},\ }\href {http://dx.doi.org/10.1038/nphoton.2013.13} {\bibfield
  {journal} {\bibinfo  {journal} {Nat. Photon.}\ }\textbf {\bibinfo {volume}
  {7}},\ \bibinfo {pages} {210} (\bibinfo {year} {2013})}\BibitemShut {NoStop}%
\bibitem [{\citenamefont {Gu}\ \emph {et~al.}(2017)\citenamefont {Gu},
  \citenamefont {Kockum}, \citenamefont {Miranowicz}, \citenamefont {Liu},\
  and\ \citenamefont {Nori}}]{Gux17phr}%
  \BibitemOpen
  \bibfield  {author} {\bibinfo {author} {\bibfnamefont {X.}~\bibnamefont
  {Gu}}, \bibinfo {author} {\bibfnamefont {A.~F.}\ \bibnamefont {Kockum}},
  \bibinfo {author} {\bibfnamefont {A.}~\bibnamefont {Miranowicz}}, \bibinfo
  {author} {\bibfnamefont {Y.}~\bibnamefont {Liu}}, \ and\ \bibinfo {author}
  {\bibfnamefont {F.}~\bibnamefont {Nori}},\ }\href {\doibase
  https://doi.org/10.1016/j.physrep.2017.10.002} {\bibfield  {journal}
  {\bibinfo  {journal} {Phys. Rep.}\ }\textbf {\bibinfo {volume} {718-719}},\
  \bibinfo {pages} {1} (\bibinfo {year} {2017})}\BibitemShut {NoStop}%
\bibitem [{\citenamefont {Devoret}\ and\ \citenamefont
  {Schoelkopf}(2013)}]{Dev13sci}%
  \BibitemOpen
  \bibfield  {author} {\bibinfo {author} {\bibfnamefont {M.~H.}\ \bibnamefont
  {Devoret}}\ and\ \bibinfo {author} {\bibfnamefont {R.~J.}\ \bibnamefont
  {Schoelkopf}},\ }\href {\doibase 10.1126/science.1231930} {\bibfield
  {journal} {\bibinfo  {journal} {Science}\ }\textbf {\bibinfo {volume}
  {339}},\ \bibinfo {pages} {1169} (\bibinfo {year} {2013})}\BibitemShut
  {NoStop}%
\bibitem [{\citenamefont {Linder}\ and\ \citenamefont
  {Robinson}(2015)}]{Lin15nph}%
  \BibitemOpen
  \bibfield  {author} {\bibinfo {author} {\bibfnamefont {J.}~\bibnamefont
  {Linder}}\ and\ \bibinfo {author} {\bibfnamefont {J.~W.~A.}\ \bibnamefont
  {Robinson}},\ }\href {http://dx.doi.org/10.1038/nphys3242} {\bibfield
  {journal} {\bibinfo  {journal} {Nat. Phys.}\ }\textbf {\bibinfo {volume}
  {11}},\ \bibinfo {pages} {307} (\bibinfo {year} {2015})}\BibitemShut
  {NoStop}%
\bibitem [{\citenamefont {Kim}\ \emph {et~al.}(2018)\citenamefont {Kim},
  \citenamefont {Myers},\ and\ \citenamefont {Tserkovnyak}}]{Kim18prl}%
  \BibitemOpen
  \bibfield  {author} {\bibinfo {author} {\bibfnamefont {S.~K.}\ \bibnamefont
  {Kim}}, \bibinfo {author} {\bibfnamefont {R.}~\bibnamefont {Myers}}, \ and\
  \bibinfo {author} {\bibfnamefont {Y.}~\bibnamefont {Tserkovnyak}},\ }\href
  {\doibase 10.1103/PhysRevLett.121.187203} {\bibfield  {journal} {\bibinfo
  {journal} {Phys. Rev. Lett.}\ }\textbf {\bibinfo {volume} {121}},\ \bibinfo
  {pages} {187203} (\bibinfo {year} {2018})}\BibitemShut {NoStop}%
\bibitem [{\citenamefont {Jeon}\ \emph {et~al.}(2018)\citenamefont {Jeon},
  \citenamefont {Ciccarelli}, \citenamefont {Ferguson}, \citenamefont
  {Kurebayashi}, \citenamefont {Cohen}, \citenamefont {Montiel}, \citenamefont
  {Eschrig}, \citenamefont {Robinson},\ and\ \citenamefont
  {Blamire}}]{Jeo18nam}%
  \BibitemOpen
  \bibfield  {author} {\bibinfo {author} {\bibfnamefont {K.-R.}\ \bibnamefont
  {Jeon}}, \bibinfo {author} {\bibfnamefont {C.}~\bibnamefont {Ciccarelli}},
  \bibinfo {author} {\bibfnamefont {A.~J.}\ \bibnamefont {Ferguson}}, \bibinfo
  {author} {\bibfnamefont {H.}~\bibnamefont {Kurebayashi}}, \bibinfo {author}
  {\bibfnamefont {L.~F.}\ \bibnamefont {Cohen}}, \bibinfo {author}
  {\bibfnamefont {X.}~\bibnamefont {Montiel}}, \bibinfo {author} {\bibfnamefont
  {M.}~\bibnamefont {Eschrig}}, \bibinfo {author} {\bibfnamefont {J.~W.~A.}\
  \bibnamefont {Robinson}}, \ and\ \bibinfo {author} {\bibfnamefont {M.~G.}\
  \bibnamefont {Blamire}},\ }\href {\doibase 10.1038/s41563-018-0058-9}
  {\bibfield  {journal} {\bibinfo  {journal} {Nat. Mater.}\ }\textbf {\bibinfo
  {volume} {17}},\ \bibinfo {pages} {499} (\bibinfo {year} {2018})}\BibitemShut
  {NoStop}%
\bibitem [{\citenamefont {Jeon}\ \emph {et~al.}(2019)\citenamefont {Jeon},
  \citenamefont {Ciccarelli}, \citenamefont {Kurebayashi}, \citenamefont
  {Cohen}, \citenamefont {Montiel}, \citenamefont {Eschrig}, \citenamefont
  {Wagner}, \citenamefont {Komori}, \citenamefont {Srivastava}, \citenamefont
  {Robinson},\ and\ \citenamefont {Blamire}}]{Jeo19pra}%
  \BibitemOpen
  \bibfield  {author} {\bibinfo {author} {\bibfnamefont {K.-R.}\ \bibnamefont
  {Jeon}}, \bibinfo {author} {\bibfnamefont {C.}~\bibnamefont {Ciccarelli}},
  \bibinfo {author} {\bibfnamefont {H.}~\bibnamefont {Kurebayashi}}, \bibinfo
  {author} {\bibfnamefont {L.~F.}\ \bibnamefont {Cohen}}, \bibinfo {author}
  {\bibfnamefont {X.}~\bibnamefont {Montiel}}, \bibinfo {author} {\bibfnamefont
  {M.}~\bibnamefont {Eschrig}}, \bibinfo {author} {\bibfnamefont
  {T.}~\bibnamefont {Wagner}}, \bibinfo {author} {\bibfnamefont
  {S.}~\bibnamefont {Komori}}, \bibinfo {author} {\bibfnamefont
  {A.}~\bibnamefont {Srivastava}}, \bibinfo {author} {\bibfnamefont {J.~W.~.}\
  \bibnamefont {Robinson}}, \ and\ \bibinfo {author} {\bibfnamefont {M.~G.}\
  \bibnamefont {Blamire}},\ }\href {\doibase 10.1103/PhysRevApplied.11.014061}
  {\bibfield  {journal} {\bibinfo  {journal} {Phys. Rev. Appl.}\ }\textbf
  {\bibinfo {volume} {11}},\ \bibinfo {pages} {014061} (\bibinfo {year}
  {2019})}\BibitemShut {NoStop}%
\bibitem [{\citenamefont {Golovchanskiy}\ \emph
  {et~al.}(2018{\natexlab{a}})\citenamefont {Golovchanskiy}, \citenamefont
  {Abramov}, \citenamefont {Stolyarov}, \citenamefont {Bolginov}, \citenamefont
  {Ryazanov}, \citenamefont {Golubov},\ and\ \citenamefont
  {Ustinov}}]{Gol18afm}%
  \BibitemOpen
  \bibfield  {author} {\bibinfo {author} {\bibfnamefont {I.~A.}\ \bibnamefont
  {Golovchanskiy}}, \bibinfo {author} {\bibfnamefont {N.~N.}\ \bibnamefont
  {Abramov}}, \bibinfo {author} {\bibfnamefont {V.~S.}\ \bibnamefont
  {Stolyarov}}, \bibinfo {author} {\bibfnamefont {V.~V.}\ \bibnamefont
  {Bolginov}}, \bibinfo {author} {\bibfnamefont {V.~V.}\ \bibnamefont
  {Ryazanov}}, \bibinfo {author} {\bibfnamefont {A.~A.}\ \bibnamefont
  {Golubov}}, \ and\ \bibinfo {author} {\bibfnamefont {A.~V.}\ \bibnamefont
  {Ustinov}},\ }\href {\doibase 10.1002/adfm.201802375} {\bibfield  {journal}
  {\bibinfo  {journal} {Adv. Func. Mater.}\ }\textbf {\bibinfo {volume} {28}},\
  \bibinfo {pages} {1802375} (\bibinfo {year}
  {2018}{\natexlab{a}})}\BibitemShut {NoStop}%
\bibitem [{\citenamefont {Dobrovolskiy}\ \emph
  {et~al.}(2019{\natexlab{a}})\citenamefont {Dobrovolskiy}, \citenamefont
  {Sachser}, \citenamefont {Br{\"a}cher}, \citenamefont {B{\"o}ttcher},
  \citenamefont {Kruglyak}, \citenamefont {Vovk}, \citenamefont {Shklovskij},
  \citenamefont {Huth}, \citenamefont {Hillebrands},\ and\ \citenamefont
  {Chumak}}]{Dob19nph}%
  \BibitemOpen
  \bibfield  {author} {\bibinfo {author} {\bibfnamefont {O.~V.}\ \bibnamefont
  {Dobrovolskiy}}, \bibinfo {author} {\bibfnamefont {R.}~\bibnamefont
  {Sachser}}, \bibinfo {author} {\bibfnamefont {T.}~\bibnamefont
  {Br{\"a}cher}}, \bibinfo {author} {\bibfnamefont {T.}~\bibnamefont
  {B{\"o}ttcher}}, \bibinfo {author} {\bibfnamefont {V.~V.}\ \bibnamefont
  {Kruglyak}}, \bibinfo {author} {\bibfnamefont {R.~V.}\ \bibnamefont {Vovk}},
  \bibinfo {author} {\bibfnamefont {V.~A.}\ \bibnamefont {Shklovskij}},
  \bibinfo {author} {\bibfnamefont {M.}~\bibnamefont {Huth}}, \bibinfo {author}
  {\bibfnamefont {B.}~\bibnamefont {Hillebrands}}, \ and\ \bibinfo {author}
  {\bibfnamefont {A.~V.}\ \bibnamefont {Chumak}},\ }\href {\doibase
  10.1038/s41567-019-0428-5} {\bibfield  {journal} {\bibinfo  {journal} {Nat.
  Phys.}\ }\textbf {\bibinfo {volume} {15}},\ \bibinfo {pages} {477} (\bibinfo
  {year} {2019}{\natexlab{a}})}\BibitemShut {NoStop}%
\bibitem [{\citenamefont {Wang}\ \emph {et~al.}(2006)\citenamefont {Wang},
  \citenamefont {Nisoli}, \citenamefont {Freitas}, \citenamefont {Li},
  \citenamefont {Cooley}, \citenamefont {Lund}, \citenamefont {Samarth},
  \citenamefont {Leighton}, \citenamefont {Crespi},\ and\ \citenamefont
  {Schiffer}}]{Wan06nat}%
  \BibitemOpen
  \bibfield  {author} {\bibinfo {author} {\bibfnamefont {R.~F.}\ \bibnamefont
  {Wang}}, \bibinfo {author} {\bibfnamefont {C.}~\bibnamefont {Nisoli}},
  \bibinfo {author} {\bibfnamefont {R.~S.}\ \bibnamefont {Freitas}}, \bibinfo
  {author} {\bibfnamefont {J.}~\bibnamefont {Li}}, \bibinfo {author}
  {\bibfnamefont {B.~J.}\ \bibnamefont {Cooley}}, \bibinfo {author}
  {\bibfnamefont {M.~S.}\ \bibnamefont {Lund}}, \bibinfo {author}
  {\bibfnamefont {N.}~\bibnamefont {Samarth}}, \bibinfo {author} {\bibfnamefont
  {C.}~\bibnamefont {Leighton}}, \bibinfo {author} {\bibfnamefont {V.~H.}\
  \bibnamefont {Crespi}}, \ and\ \bibinfo {author} {\bibfnamefont
  {P.}~\bibnamefont {Schiffer}},\ }\href {\doibase
  http://dx.doi.org/10.1038/nature04447} {\bibfield  {journal} {\bibinfo
  {journal} {Nature}\ }\textbf {\bibinfo {volume} {439}},\ \bibinfo {pages}
  {303} (\bibinfo {year} {2006})}\BibitemShut {NoStop}%
\bibitem [{\citenamefont {Bespalov}\ \emph {et~al.}(2014)\citenamefont
  {Bespalov}, \citenamefont {Mel'nikov},\ and\ \citenamefont
  {Buzdin}}]{Bes14prb}%
  \BibitemOpen
  \bibfield  {author} {\bibinfo {author} {\bibfnamefont {A.~A.}\ \bibnamefont
  {Bespalov}}, \bibinfo {author} {\bibfnamefont {A.~S.}\ \bibnamefont
  {Mel'nikov}}, \ and\ \bibinfo {author} {\bibfnamefont {A.~I.}\ \bibnamefont
  {Buzdin}},\ }\href {\doibase 10.1103/PhysRevB.89.054516} {\bibfield
  {journal} {\bibinfo  {journal} {Phys. Rev. B}\ }\textbf {\bibinfo {volume}
  {89}},\ \bibinfo {pages} {054516} (\bibinfo {year} {2014})}\BibitemShut
  {NoStop}%
\bibitem [{\citenamefont {Stolyarov}\ \emph {et~al.}(2018)\citenamefont
  {Stolyarov}, \citenamefont {Veshchunov}, \citenamefont {Grebenchuk},
  \citenamefont {Baranov}, \citenamefont {Golovchanskiy}, \citenamefont
  {Shishkin}, \citenamefont {Zhou}, \citenamefont {Shi}, \citenamefont {Xu},
  \citenamefont {Pyon}, \citenamefont {Sun}, \citenamefont {Jiao},
  \citenamefont {Cao}, \citenamefont {Vinnikov}, \citenamefont {Tamegai},
  \citenamefont {Buzdin},\ and\ \citenamefont {Roditchev}}]{Sto18adv}%
  \BibitemOpen
  \bibfield  {author} {\bibinfo {author} {\bibfnamefont {V.~S.}\ \bibnamefont
  {Stolyarov}}, \bibinfo {author} {\bibfnamefont {I.~S.}\ \bibnamefont
  {Veshchunov}}, \bibinfo {author} {\bibfnamefont {S.~Y.}\ \bibnamefont
  {Grebenchuk}}, \bibinfo {author} {\bibfnamefont {D.~S.}\ \bibnamefont
  {Baranov}}, \bibinfo {author} {\bibfnamefont {I.~A.}\ \bibnamefont
  {Golovchanskiy}}, \bibinfo {author} {\bibfnamefont {A.~G.}\ \bibnamefont
  {Shishkin}}, \bibinfo {author} {\bibfnamefont {N.}~\bibnamefont {Zhou}},
  \bibinfo {author} {\bibfnamefont {Z.}~\bibnamefont {Shi}}, \bibinfo {author}
  {\bibfnamefont {X.}~\bibnamefont {Xu}}, \bibinfo {author} {\bibfnamefont
  {S.}~\bibnamefont {Pyon}}, \bibinfo {author} {\bibfnamefont {Y.}~\bibnamefont
  {Sun}}, \bibinfo {author} {\bibfnamefont {W.}~\bibnamefont {Jiao}}, \bibinfo
  {author} {\bibfnamefont {G.-H.}\ \bibnamefont {Cao}}, \bibinfo {author}
  {\bibfnamefont {A.~A.}\ \bibnamefont {Vinnikov}, \bibfnamefont {L.~Ya
  .and~Golubov}}, \bibinfo {author} {\bibfnamefont {T.}~\bibnamefont
  {Tamegai}}, \bibinfo {author} {\bibfnamefont {A.~I.}\ \bibnamefont {Buzdin}},
  \ and\ \bibinfo {author} {\bibfnamefont {D.}~\bibnamefont {Roditchev}},\
  }\href {\doibase 10.1126/sciadv.aat1061} {\bibfield  {journal} {\bibinfo
  {journal} {Sci. Adv.}\ }\textbf {\bibinfo {volume} {4}},\ \bibinfo {pages}
  {eaat1061} (\bibinfo {year} {2018})}\BibitemShut {NoStop}%
\bibitem [{\citenamefont {Embon}\ \emph {et~al.}(2017)\citenamefont {Embon},
  \citenamefont {Anahory}, \citenamefont {Jelic}, \citenamefont {Lachman},
  \citenamefont {Myasoedov}, \citenamefont {Huber}, \citenamefont {Mikitik},
  \citenamefont {Silhanek}, \citenamefont {Milosevic}, \citenamefont
  {Gurevich},\ and\ \citenamefont {Zeldov}}]{Emb17nac}%
  \BibitemOpen
  \bibfield  {author} {\bibinfo {author} {\bibfnamefont {L.}~\bibnamefont
  {Embon}}, \bibinfo {author} {\bibfnamefont {Y.}~\bibnamefont {Anahory}},
  \bibinfo {author} {\bibfnamefont {Z.~L.}\ \bibnamefont {Jelic}}, \bibinfo
  {author} {\bibfnamefont {E.~O.}\ \bibnamefont {Lachman}}, \bibinfo {author}
  {\bibfnamefont {Y.}~\bibnamefont {Myasoedov}}, \bibinfo {author}
  {\bibfnamefont {M.~E.}\ \bibnamefont {Huber}}, \bibinfo {author}
  {\bibfnamefont {G.~P.}\ \bibnamefont {Mikitik}}, \bibinfo {author}
  {\bibfnamefont {A.~V.}\ \bibnamefont {Silhanek}}, \bibinfo {author}
  {\bibfnamefont {M.~V.}\ \bibnamefont {Milosevic}}, \bibinfo {author}
  {\bibfnamefont {A.}~\bibnamefont {Gurevich}}, \ and\ \bibinfo {author}
  {\bibfnamefont {E.}~\bibnamefont {Zeldov}},\ }\href {\doibase
  10.1038/s41467-017-00089-3} {\bibfield  {journal} {\bibinfo  {journal} {Nat.
  Comms.}\ }\textbf {\bibinfo {volume} {8}},\ \bibinfo {pages} {85} (\bibinfo
  {year} {2017})}\BibitemShut {NoStop}%
\bibitem [{\citenamefont {Yue}\ \emph {et~al.}(2017)\citenamefont {Yue},
  \citenamefont {Chen}, \citenamefont {Barreda}, \citenamefont {Bevara},
  \citenamefont {Hu}, \citenamefont {Wu}, \citenamefont {Wang}, \citenamefont
  {Andrei}, \citenamefont {Bertaina},\ and\ \citenamefont
  {Chiorescu}}]{Yue17apl}%
  \BibitemOpen
  \bibfield  {author} {\bibinfo {author} {\bibfnamefont {G.}~\bibnamefont
  {Yue}}, \bibinfo {author} {\bibfnamefont {L.}~\bibnamefont {Chen}}, \bibinfo
  {author} {\bibfnamefont {J.}~\bibnamefont {Barreda}}, \bibinfo {author}
  {\bibfnamefont {V.}~\bibnamefont {Bevara}}, \bibinfo {author} {\bibfnamefont
  {L.}~\bibnamefont {Hu}}, \bibinfo {author} {\bibfnamefont {L.}~\bibnamefont
  {Wu}}, \bibinfo {author} {\bibfnamefont {Z.}~\bibnamefont {Wang}}, \bibinfo
  {author} {\bibfnamefont {P.}~\bibnamefont {Andrei}}, \bibinfo {author}
  {\bibfnamefont {S.}~\bibnamefont {Bertaina}}, \ and\ \bibinfo {author}
  {\bibfnamefont {I.}~\bibnamefont {Chiorescu}},\ }\href {\doibase
  10.1063/1.5006693} {\bibfield  {journal} {\bibinfo  {journal} {Appl. Phys.
  Lett.}\ }\textbf {\bibinfo {volume} {111}},\ \bibinfo {pages} {202601}
  (\bibinfo {year} {2017})}\BibitemShut {NoStop}%
\bibitem [{\citenamefont {Golovchanskiy}\ \emph
  {et~al.}(2018{\natexlab{b}})\citenamefont {Golovchanskiy}, \citenamefont
  {Abramov}, \citenamefont {Stolyarov}, \citenamefont {Shchetinin},
  \citenamefont {Dzhumaev}, \citenamefont {Averkin}, \citenamefont {Kozlov},
  \citenamefont {Golubov}, \citenamefont {Ryazanov},\ and\ \citenamefont
  {Ustinov}}]{Gol18jap}%
  \BibitemOpen
  \bibfield  {author} {\bibinfo {author} {\bibfnamefont {I.~A.}\ \bibnamefont
  {Golovchanskiy}}, \bibinfo {author} {\bibfnamefont {N.~N.}\ \bibnamefont
  {Abramov}}, \bibinfo {author} {\bibfnamefont {V.~S.}\ \bibnamefont
  {Stolyarov}}, \bibinfo {author} {\bibfnamefont {I.~V.}\ \bibnamefont
  {Shchetinin}}, \bibinfo {author} {\bibfnamefont {P.~S.}\ \bibnamefont
  {Dzhumaev}}, \bibinfo {author} {\bibfnamefont {A.~S.}\ \bibnamefont
  {Averkin}}, \bibinfo {author} {\bibfnamefont {S.~N.}\ \bibnamefont {Kozlov}},
  \bibinfo {author} {\bibfnamefont {A.~A.}\ \bibnamefont {Golubov}}, \bibinfo
  {author} {\bibfnamefont {V.~V.}\ \bibnamefont {Ryazanov}}, \ and\ \bibinfo
  {author} {\bibfnamefont {A.~V.}\ \bibnamefont {Ustinov}},\ }\href {\doibase
  10.1063/1.5025028} {\bibfield  {journal} {\bibinfo  {journal} {J. Appl.
  Phys.}\ }\textbf {\bibinfo {volume} {123}},\ \bibinfo {pages} {173904}
  (\bibinfo {year} {2018}{\natexlab{b}})}\BibitemShut {NoStop}%
\bibitem [{\citenamefont {Girvin}(2014)}]{Gir14inp}%
  \BibitemOpen
  \bibfield  {author} {\bibinfo {author} {\bibfnamefont {S.~M.}\ \bibnamefont
  {Girvin}},\ }in\ \href@noop {} {\emph {\bibinfo {booktitle} {Proc. of the
  2011 Les Houches Summer School on Quantum Machines}}}\ (\bibinfo {year}
  {2014})\BibitemShut {NoStop}%
\bibitem [{\citenamefont {Larkin}\ and\ \citenamefont
  {Ovchinnikov}(1975)}]{Lar75etp}%
  \BibitemOpen
  \bibfield  {author} {\bibinfo {author} {\bibfnamefont {A.~I.}\ \bibnamefont
  {Larkin}}\ and\ \bibinfo {author} {\bibfnamefont {Y.~N.}\ \bibnamefont
  {Ovchinnikov}},\ }\href
  {http://www.jetp.ac.ru/cgi-bin/index/e/41/5/p960?a=list} {\bibfield
  {journal} {\bibinfo  {journal} {J. Exp. Theor. Phys.}\ }\textbf {\bibinfo
  {volume} {41}},\ \bibinfo {pages} {960} (\bibinfo {year} {1975})}\BibitemShut
  {NoStop}%
\bibitem [{\citenamefont {Larkin}\ and\ \citenamefont
  {Ovchinnikov}(1986)}]{Lar86inb}%
  \BibitemOpen
  \bibfield  {author} {\bibinfo {author} {\bibfnamefont {A.~I.}\ \bibnamefont
  {Larkin}}\ and\ \bibinfo {author} {\bibfnamefont {Y.~N.}\ \bibnamefont
  {Ovchinnikov}},\ }\enquote {\bibinfo {title} {Nonequilibrium
  superconductivity},}\ \ (\bibinfo  {publisher} {Elsevier, Amsterdam},\
  \bibinfo {year} {1986})\ p.\ \bibinfo {pages} {493}\BibitemShut {NoStop}%
\bibitem [{\citenamefont {Bezuglyj}\ and\ \citenamefont
  {Shklovskij}(1992)}]{Bez92pcs}%
  \BibitemOpen
  \bibfield  {author} {\bibinfo {author} {\bibfnamefont {A.}~\bibnamefont
  {Bezuglyj}}\ and\ \bibinfo {author} {\bibfnamefont {V.}~\bibnamefont
  {Shklovskij}},\ }\href {\doibase 10.1016/0921-4534(92)90165-9} {\bibfield
  {journal} {\bibinfo  {journal} {Physica C}\ }\textbf {\bibinfo {volume}
  {202}},\ \bibinfo {pages} {234} (\bibinfo {year} {1992})}\BibitemShut
  {NoStop}%
\bibitem [{\citenamefont {Gray}(1981)}]{Gra81boo}%
  \BibitemOpen
  \bibinfo {editor} {\bibfnamefont {K.~E.}\ \bibnamefont {Gray}},\ ed.,\
  \href@noop {} {\emph {\bibinfo {title} {Nonequilibrium Superconductivity,
  Phonons, and Kapitza Boundaries}}}\ (\bibinfo  {publisher} {Plenum press, New
  York and London},\ \bibinfo {year} {1981})\BibitemShut {NoStop}%
\bibitem [{\citenamefont {Kopnin}(2001)}]{Kop01boo}%
  \BibitemOpen
  \bibfield  {author} {\bibinfo {author} {\bibfnamefont {N.~B.}\ \bibnamefont
  {Kopnin}},\ }\href@noop {} {\emph {\bibinfo {title} {Theory of Nonequilibrium
  Superconductivity}}}\ (\bibinfo  {publisher} {Oxford University Press, New
  York},\ \bibinfo {year} {2001})\BibitemShut {NoStop}%
\bibitem [{\citenamefont {Serniak}\ \emph {et~al.}(2018)\citenamefont
  {Serniak}, \citenamefont {Hays}, \citenamefont {de~Lange}, \citenamefont
  {Diamond}, \citenamefont {Shankar}, \citenamefont {Burkhart}, \citenamefont
  {Frunzio}, \citenamefont {Houzet},\ and\ \citenamefont {Devoret}}]{Ser18prl}%
  \BibitemOpen
  \bibfield  {author} {\bibinfo {author} {\bibfnamefont {K.}~\bibnamefont
  {Serniak}}, \bibinfo {author} {\bibfnamefont {M.}~\bibnamefont {Hays}},
  \bibinfo {author} {\bibfnamefont {G.}~\bibnamefont {de~Lange}}, \bibinfo
  {author} {\bibfnamefont {S.}~\bibnamefont {Diamond}}, \bibinfo {author}
  {\bibfnamefont {S.}~\bibnamefont {Shankar}}, \bibinfo {author} {\bibfnamefont
  {L.~D.}\ \bibnamefont {Burkhart}}, \bibinfo {author} {\bibfnamefont
  {L.}~\bibnamefont {Frunzio}}, \bibinfo {author} {\bibfnamefont
  {M.}~\bibnamefont {Houzet}}, \ and\ \bibinfo {author} {\bibfnamefont {M.~H.}\
  \bibnamefont {Devoret}},\ }\href {\doibase 10.1103/PhysRevLett.121.157701}
  {\bibfield  {journal} {\bibinfo  {journal} {Phys. Rev. Lett.}\ }\textbf
  {\bibinfo {volume} {121}},\ \bibinfo {pages} {157701} (\bibinfo {year}
  {2018})}\BibitemShut {NoStop}%
\bibitem [{\citenamefont {Eliashberg}(1970)}]{Eli70etp}%
  \BibitemOpen
  \bibfield  {author} {\bibinfo {author} {\bibfnamefont {G.~M.}\ \bibnamefont
  {Eliashberg}},\ }\href
  {http://www.jetpletters.ac.ru/ps/1716/article\_26086.pdf} {\bibfield
  {journal} {\bibinfo  {journal} {JETP Lett.}\ }\textbf {\bibinfo {volume}
  {11}},\ \bibinfo {pages} {186} (\bibinfo {year} {1970})}\BibitemShut
  {NoStop}%
\bibitem [{\citenamefont {Gurevich}\ and\ \citenamefont
  {Ciovati}(2008)}]{Gur08prb}%
  \BibitemOpen
  \bibfield  {author} {\bibinfo {author} {\bibfnamefont {A.}~\bibnamefont
  {Gurevich}}\ and\ \bibinfo {author} {\bibfnamefont {G.}~\bibnamefont
  {Ciovati}},\ }\href {\doibase 10.1103/PhysRevB.77.104501} {\bibfield
  {journal} {\bibinfo  {journal} {Phys. Rev. B}\ }\textbf {\bibinfo {volume}
  {77}},\ \bibinfo {pages} {104501} (\bibinfo {year} {2008})}\BibitemShut
  {NoStop}%
\bibitem [{\citenamefont {Pompeo}\ and\ \citenamefont
  {Silva}(2008)}]{Pom08prb}%
  \BibitemOpen
  \bibfield  {author} {\bibinfo {author} {\bibfnamefont {N.}~\bibnamefont
  {Pompeo}}\ and\ \bibinfo {author} {\bibfnamefont {E.}~\bibnamefont {Silva}},\
  }\href {\doibase 10.1103/PhysRevB.78.094503} {\bibfield  {journal} {\bibinfo
  {journal} {Phys. Rev. B}\ }\textbf {\bibinfo {volume} {78}},\ \bibinfo
  {pages} {094503} (\bibinfo {year} {2008})}\BibitemShut {NoStop}%
\bibitem [{\citenamefont {Kogan}(2018)}]{Kog18prb}%
  \BibitemOpen
  \bibfield  {author} {\bibinfo {author} {\bibfnamefont {V.~G.}\ \bibnamefont
  {Kogan}},\ }\href {\doibase 10.1103/PhysRevB.97.094510} {\bibfield  {journal}
  {\bibinfo  {journal} {Phys. Rev. B}\ }\textbf {\bibinfo {volume} {97}},\
  \bibinfo {pages} {094510} (\bibinfo {year} {2018})}\BibitemShut {NoStop}%
\bibitem [{\citenamefont {Wyatt}\ \emph {et~al.}(1966)\citenamefont {Wyatt},
  \citenamefont {Dmitriev}, \citenamefont {Moore},\ and\ \citenamefont
  {Sheard}}]{Wya66prl}%
  \BibitemOpen
  \bibfield  {author} {\bibinfo {author} {\bibfnamefont {A.~F.~G.}\
  \bibnamefont {Wyatt}}, \bibinfo {author} {\bibfnamefont {V.~M.}\ \bibnamefont
  {Dmitriev}}, \bibinfo {author} {\bibfnamefont {W.~S.}\ \bibnamefont {Moore}},
  \ and\ \bibinfo {author} {\bibfnamefont {F.~W.}\ \bibnamefont {Sheard}},\
  }\href {\doibase 10.1103/PhysRevLett.16.1166} {\bibfield  {journal} {\bibinfo
   {journal} {Phys. Rev. Lett.}\ }\textbf {\bibinfo {volume} {16}},\ \bibinfo
  {pages} {1166} (\bibinfo {year} {1966})}\BibitemShut {NoStop}%
\bibitem [{\citenamefont {Zolochevskii}(2013)}]{Zol13ltp}%
  \BibitemOpen
  \bibfield  {author} {\bibinfo {author} {\bibfnamefont {I.~V.}\ \bibnamefont
  {Zolochevskii}},\ }\href {\doibase 10.1063/1.4813655} {\bibfield  {journal}
  {\bibinfo  {journal} {Low Temp. Phys.}\ }\textbf {\bibinfo {volume} {39}},\
  \bibinfo {pages} {571} (\bibinfo {year} {2013})}\BibitemShut {NoStop}%
\bibitem [{\citenamefont {Beck}\ \emph {et~al.}(2013)\citenamefont {Beck},
  \citenamefont {Rousseau}, \citenamefont {Klammer}, \citenamefont {Leiderer},
  \citenamefont {Mittendorff}, \citenamefont {Winnerl}, \citenamefont {Helm},
  \citenamefont {Gol'tsman},\ and\ \citenamefont {Demsar}}]{Bec13prl}%
  \BibitemOpen
  \bibfield  {author} {\bibinfo {author} {\bibfnamefont {M.}~\bibnamefont
  {Beck}}, \bibinfo {author} {\bibfnamefont {I.}~\bibnamefont {Rousseau}},
  \bibinfo {author} {\bibfnamefont {M.}~\bibnamefont {Klammer}}, \bibinfo
  {author} {\bibfnamefont {P.}~\bibnamefont {Leiderer}}, \bibinfo {author}
  {\bibfnamefont {M.}~\bibnamefont {Mittendorff}}, \bibinfo {author}
  {\bibfnamefont {S.}~\bibnamefont {Winnerl}}, \bibinfo {author} {\bibfnamefont
  {M.}~\bibnamefont {Helm}}, \bibinfo {author} {\bibfnamefont {G.~N.}\
  \bibnamefont {Gol'tsman}}, \ and\ \bibinfo {author} {\bibfnamefont
  {J.}~\bibnamefont {Demsar}},\ }\href {\doibase
  10.1103/PhysRevLett.110.267003} {\bibfield  {journal} {\bibinfo  {journal}
  {Phys. Rev. Lett.}\ }\textbf {\bibinfo {volume} {110}},\ \bibinfo {pages}
  {267003} (\bibinfo {year} {2013})}\BibitemShut {NoStop}%
\bibitem [{\citenamefont {de~Visser}\ \emph {et~al.}(2014)\citenamefont
  {de~Visser}, \citenamefont {Goldie}, \citenamefont {Diener}, \citenamefont
  {Withington}, \citenamefont {Baselmans},\ and\ \citenamefont
  {Klapwijk}}]{Vis14prl}%
  \BibitemOpen
  \bibfield  {author} {\bibinfo {author} {\bibfnamefont {P.~J.}\ \bibnamefont
  {de~Visser}}, \bibinfo {author} {\bibfnamefont {D.~J.}\ \bibnamefont
  {Goldie}}, \bibinfo {author} {\bibfnamefont {P.}~\bibnamefont {Diener}},
  \bibinfo {author} {\bibfnamefont {S.}~\bibnamefont {Withington}}, \bibinfo
  {author} {\bibfnamefont {J.~J.~A.}\ \bibnamefont {Baselmans}}, \ and\
  \bibinfo {author} {\bibfnamefont {T.~M.}\ \bibnamefont {Klapwijk}},\ }\href
  {\doibase 10.1103/PhysRevLett.112.047004} {\bibfield  {journal} {\bibinfo
  {journal} {Phys. Rev. Lett.}\ }\textbf {\bibinfo {volume} {112}},\ \bibinfo
  {pages} {047004} (\bibinfo {year} {2014})}\BibitemShut {NoStop}%
\bibitem [{\citenamefont {Hartnoll}\ \emph {et~al.}(2008)\citenamefont
  {Hartnoll}, \citenamefont {Herzog},\ and\ \citenamefont
  {Horowitz}}]{Har08prl}%
  \BibitemOpen
  \bibfield  {author} {\bibinfo {author} {\bibfnamefont {S.~A.}\ \bibnamefont
  {Hartnoll}}, \bibinfo {author} {\bibfnamefont {C.~P.}\ \bibnamefont
  {Herzog}}, \ and\ \bibinfo {author} {\bibfnamefont {G.~T.}\ \bibnamefont
  {Horowitz}},\ }\href {\doibase 10.1103/PhysRevLett.101.031601} {\bibfield
  {journal} {\bibinfo  {journal} {Phys. Rev. Lett.}\ }\textbf {\bibinfo
  {volume} {101}},\ \bibinfo {pages} {031601} (\bibinfo {year}
  {2008})}\BibitemShut {NoStop}%
\bibitem [{\citenamefont {Montull}\ \emph {et~al.}(2009)\citenamefont
  {Montull}, \citenamefont {Pomarol},\ and\ \citenamefont {Silva}}]{Mon09prl}%
  \BibitemOpen
  \bibfield  {author} {\bibinfo {author} {\bibfnamefont {M.}~\bibnamefont
  {Montull}}, \bibinfo {author} {\bibfnamefont {A.}~\bibnamefont {Pomarol}}, \
  and\ \bibinfo {author} {\bibfnamefont {P.~J.}\ \bibnamefont {Silva}},\ }\href
  {\doibase 10.1103/PhysRevLett.103.091601} {\bibfield  {journal} {\bibinfo
  {journal} {Phys. Rev. Lett.}\ }\textbf {\bibinfo {volume} {103}},\ \bibinfo
  {pages} {091601} (\bibinfo {year} {2009})}\BibitemShut {NoStop}%
\bibitem [{\citenamefont {Maeda}\ and\ \citenamefont
  {Okamura}(2011)}]{Mae11prd}%
  \BibitemOpen
  \bibfield  {author} {\bibinfo {author} {\bibfnamefont {K.}~\bibnamefont
  {Maeda}}\ and\ \bibinfo {author} {\bibfnamefont {T.}~\bibnamefont
  {Okamura}},\ }\href {\doibase 10.1103/PhysRevD.83.066004} {\bibfield
  {journal} {\bibinfo  {journal} {Phys. Rev. D}\ }\textbf {\bibinfo {volume}
  {83}},\ \bibinfo {pages} {066004} (\bibinfo {year} {2011})}\BibitemShut
  {NoStop}%
\bibitem [{\citenamefont {Bao}\ \emph {et~al.}(2011)\citenamefont {Bao},
  \citenamefont {Dong}, \citenamefont {Silverstein},\ and\ \citenamefont
  {Torroba}}]{Bao11hep}%
  \BibitemOpen
  \bibfield  {author} {\bibinfo {author} {\bibfnamefont {N.}~\bibnamefont
  {Bao}}, \bibinfo {author} {\bibfnamefont {X.}~\bibnamefont {Dong}}, \bibinfo
  {author} {\bibfnamefont {E.}~\bibnamefont {Silverstein}}, \ and\ \bibinfo
  {author} {\bibfnamefont {G.}~\bibnamefont {Torroba}},\ }\href {\doibase
  10.1007/JHEP10(2011)123} {\bibfield  {journal} {\bibinfo  {journal} {J. High
  Ener. Phys.}\ }\textbf {\bibinfo {volume} {2011}},\ \bibinfo {pages} {123}
  (\bibinfo {year} {2011})}\BibitemShut {NoStop}%
\bibitem [{\citenamefont {Natsuume}\ and\ \citenamefont
  {Okamura}(2013)}]{Nat13hep}%
  \BibitemOpen
  \bibfield  {author} {\bibinfo {author} {\bibfnamefont {M.}~\bibnamefont
  {Natsuume}}\ and\ \bibinfo {author} {\bibfnamefont {T.}~\bibnamefont
  {Okamura}},\ }\href {\doibase 10.1007/JHEP08(2013)139} {\bibfield  {journal}
  {\bibinfo  {journal} {J. High Ener. Phys.}\ }\textbf {\bibinfo {volume}
  {2013}},\ \bibinfo {pages} {139} (\bibinfo {year} {2013})}\BibitemShut
  {NoStop}%
\bibitem [{\citenamefont {Clem}(1968)}]{Cle68prl}%
  \BibitemOpen
  \bibfield  {author} {\bibinfo {author} {\bibfnamefont {J.~R.}\ \bibnamefont
  {Clem}},\ }\href {\doibase 10.1103/PhysRevLett.20.735} {\bibfield  {journal}
  {\bibinfo  {journal} {Phys. Rev. Lett.}\ }\textbf {\bibinfo {volume} {20}},\
  \bibinfo {pages} {735} (\bibinfo {year} {1968})}\BibitemShut {NoStop}%
\bibitem [{\citenamefont {Shekhter}\ \emph {et~al.}(2011)\citenamefont
  {Shekhter}, \citenamefont {Bulaevskii},\ and\ \citenamefont
  {Batista}}]{She11prl}%
  \BibitemOpen
  \bibfield  {author} {\bibinfo {author} {\bibfnamefont {A.}~\bibnamefont
  {Shekhter}}, \bibinfo {author} {\bibfnamefont {L.~N.}\ \bibnamefont
  {Bulaevskii}}, \ and\ \bibinfo {author} {\bibfnamefont {C.~D.}\ \bibnamefont
  {Batista}},\ }\href {\doibase 10.1103/PhysRevLett.106.037001} {\bibfield
  {journal} {\bibinfo  {journal} {Phys. Rev. Lett.}\ }\textbf {\bibinfo
  {volume} {106}},\ \bibinfo {pages} {037001} (\bibinfo {year}
  {2011})}\BibitemShut {NoStop}%
\bibitem [{\citenamefont {Semenov}\ \emph {et~al.}(2016)\citenamefont
  {Semenov}, \citenamefont {Devyatov}, \citenamefont {de~Visser},\ and\
  \citenamefont {Klapwijk}}]{Sem16prl}%
  \BibitemOpen
  \bibfield  {author} {\bibinfo {author} {\bibfnamefont {A.~V.}\ \bibnamefont
  {Semenov}}, \bibinfo {author} {\bibfnamefont {I.~A.}\ \bibnamefont
  {Devyatov}}, \bibinfo {author} {\bibfnamefont {P.~J.}\ \bibnamefont
  {de~Visser}}, \ and\ \bibinfo {author} {\bibfnamefont {T.~M.}\ \bibnamefont
  {Klapwijk}},\ }\href {\doibase 10.1103/PhysRevLett.117.047002} {\bibfield
  {journal} {\bibinfo  {journal} {Phys. Rev. Lett.}\ }\textbf {\bibinfo
  {volume} {117}},\ \bibinfo {pages} {047002} (\bibinfo {year}
  {2016})}\BibitemShut {NoStop}%
\bibitem [{\citenamefont {Tikhonov}\ \emph {et~al.}(2018)\citenamefont
  {Tikhonov}, \citenamefont {Skvortsov},\ and\ \citenamefont
  {Klapwijk}}]{Tik18prb}%
  \BibitemOpen
  \bibfield  {author} {\bibinfo {author} {\bibfnamefont {K.~S.}\ \bibnamefont
  {Tikhonov}}, \bibinfo {author} {\bibfnamefont {M.~A.}\ \bibnamefont
  {Skvortsov}}, \ and\ \bibinfo {author} {\bibfnamefont {T.~M.}\ \bibnamefont
  {Klapwijk}},\ }\href {\doibase 10.1103/PhysRevB.97.184516} {\bibfield
  {journal} {\bibinfo  {journal} {Phys. Rev. B}\ }\textbf {\bibinfo {volume}
  {97}},\ \bibinfo {pages} {184516} (\bibinfo {year} {2018})}\BibitemShut
  {NoStop}%
\bibitem [{\citenamefont {Shklovskij}\ \emph {et~al.}(2017)\citenamefont
  {Shklovskij}, \citenamefont {Nazipova},\ and\ \citenamefont
  {Dobrovolskiy}}]{Shk17prb}%
  \BibitemOpen
  \bibfield  {author} {\bibinfo {author} {\bibfnamefont {V.~A.}\ \bibnamefont
  {Shklovskij}}, \bibinfo {author} {\bibfnamefont {A.~P.}\ \bibnamefont
  {Nazipova}}, \ and\ \bibinfo {author} {\bibfnamefont {O.~V.}\ \bibnamefont
  {Dobrovolskiy}},\ }\href {\doibase 10.1103/PhysRevB.95.184517} {\bibfield
  {journal} {\bibinfo  {journal} {Phys. Rev. B}\ }\textbf {\bibinfo {volume}
  {95}},\ \bibinfo {pages} {184517} (\bibinfo {year} {2017})}\BibitemShut
  {NoStop}%
\bibitem [{\citenamefont {Vodolazov}(2017)}]{Vod17pra}%
  \BibitemOpen
  \bibfield  {author} {\bibinfo {author} {\bibfnamefont {D.~Y.}\ \bibnamefont
  {Vodolazov}},\ }\href {\doibase 10.1103/PhysRevApplied.7.034014} {\bibfield
  {journal} {\bibinfo  {journal} {Phys. Rev. Appl.}\ }\textbf {\bibinfo
  {volume} {7}},\ \bibinfo {pages} {034014} (\bibinfo {year}
  {2017})}\BibitemShut {NoStop}%
\bibitem [{\citenamefont {Yang}\ and\ \citenamefont {Wu}(2018)}]{Yan18prb}%
  \BibitemOpen
  \bibfield  {author} {\bibinfo {author} {\bibfnamefont {F.}~\bibnamefont
  {Yang}}\ and\ \bibinfo {author} {\bibfnamefont {M.~W.}\ \bibnamefont {Wu}},\
  }\href {\doibase 10.1103/PhysRevB.98.094507} {\bibfield  {journal} {\bibinfo
  {journal} {Phys. Rev. B}\ }\textbf {\bibinfo {volume} {98}},\ \bibinfo
  {pages} {094507} (\bibinfo {year} {2018})}\BibitemShut {NoStop}%
\bibitem [{\citenamefont {Leo}\ \emph {et~al.}(2011)\citenamefont {Leo},
  \citenamefont {Grimaldi}, \citenamefont {Citro}, \citenamefont {Nigro},
  \citenamefont {Pace},\ and\ \citenamefont {Huebener}}]{Leo11prb}%
  \BibitemOpen
  \bibfield  {author} {\bibinfo {author} {\bibfnamefont {A.}~\bibnamefont
  {Leo}}, \bibinfo {author} {\bibfnamefont {G.}~\bibnamefont {Grimaldi}},
  \bibinfo {author} {\bibfnamefont {R.}~\bibnamefont {Citro}}, \bibinfo
  {author} {\bibfnamefont {A.}~\bibnamefont {Nigro}}, \bibinfo {author}
  {\bibfnamefont {S.}~\bibnamefont {Pace}}, \ and\ \bibinfo {author}
  {\bibfnamefont {R.~P.}\ \bibnamefont {Huebener}},\ }\href {\doibase
  10.1103/PhysRevB.84.014536} {\bibfield  {journal} {\bibinfo  {journal} {Phys.
  Rev. B}\ }\textbf {\bibinfo {volume} {84}},\ \bibinfo {pages} {014536}
  (\bibinfo {year} {2011})}\BibitemShut {NoStop}%
\bibitem [{\citenamefont {Silhanek}\ \emph {et~al.}(2012)\citenamefont
  {Silhanek}, \citenamefont {Leo}, \citenamefont {Grimaldi}, \citenamefont
  {Berdiyorov}, \citenamefont {Milosevic}, \citenamefont {Nigro}, \citenamefont
  {Pace}, \citenamefont {Verellen}, \citenamefont {Gillijns}, \citenamefont
  {Metlushko}, \citenamefont {Ili}, \citenamefont {Zhu},\ and\ \citenamefont
  {Moshchalkov}}]{Sil12njp}%
  \BibitemOpen
  \bibfield  {author} {\bibinfo {author} {\bibfnamefont {A.~V.}\ \bibnamefont
  {Silhanek}}, \bibinfo {author} {\bibfnamefont {A.}~\bibnamefont {Leo}},
  \bibinfo {author} {\bibfnamefont {G.}~\bibnamefont {Grimaldi}}, \bibinfo
  {author} {\bibfnamefont {G.~R.}\ \bibnamefont {Berdiyorov}}, \bibinfo
  {author} {\bibfnamefont {M.~V.}\ \bibnamefont {Milosevic}}, \bibinfo {author}
  {\bibfnamefont {A.}~\bibnamefont {Nigro}}, \bibinfo {author} {\bibfnamefont
  {S.}~\bibnamefont {Pace}}, \bibinfo {author} {\bibfnamefont {N.}~\bibnamefont
  {Verellen}}, \bibinfo {author} {\bibfnamefont {W.}~\bibnamefont {Gillijns}},
  \bibinfo {author} {\bibfnamefont {V.}~\bibnamefont {Metlushko}}, \bibinfo
  {author} {\bibfnamefont {B.}~\bibnamefont {Ili}}, \bibinfo {author}
  {\bibfnamefont {X.}~\bibnamefont {Zhu}}, \ and\ \bibinfo {author}
  {\bibfnamefont {V.~V.}\ \bibnamefont {Moshchalkov}},\ }\href
  {http://stacks.iop.org/1367-2630/14/i=5/a=053006} {\bibfield  {journal}
  {\bibinfo  {journal} {New J. Phys.}\ }\textbf {\bibinfo {volume} {14}},\
  \bibinfo {pages} {053006} (\bibinfo {year} {2012})}\BibitemShut {NoStop}%
\bibitem [{\citenamefont {Peroz}\ and\ \citenamefont
  {Villard}(2005)}]{Per05prb}%
  \BibitemOpen
  \bibfield  {author} {\bibinfo {author} {\bibfnamefont {C.}~\bibnamefont
  {Peroz}}\ and\ \bibinfo {author} {\bibfnamefont {C.}~\bibnamefont
  {Villard}},\ }\href {\doibase 10.1103/PhysRevB.72.014515} {\bibfield
  {journal} {\bibinfo  {journal} {Phys. Rev. B}\ }\textbf {\bibinfo {volume}
  {72}},\ \bibinfo {pages} {014515} (\bibinfo {year} {2005})}\BibitemShut
  {NoStop}%
\bibitem [{\citenamefont {W\"ordenweber}\ \emph {et~al.}(2012)\citenamefont
  {W\"ordenweber}, \citenamefont {Hollmann}, \citenamefont {Schubert},
  \citenamefont {Kutzner},\ and\ \citenamefont {Panaitov}}]{Wor12prb}%
  \BibitemOpen
  \bibfield  {author} {\bibinfo {author} {\bibfnamefont {R.}~\bibnamefont
  {W\"ordenweber}}, \bibinfo {author} {\bibfnamefont {E.}~\bibnamefont
  {Hollmann}}, \bibinfo {author} {\bibfnamefont {J.}~\bibnamefont {Schubert}},
  \bibinfo {author} {\bibfnamefont {R.}~\bibnamefont {Kutzner}}, \ and\
  \bibinfo {author} {\bibfnamefont {G.}~\bibnamefont {Panaitov}},\ }\href
  {\doibase 10.1103/PhysRevB.85.064503} {\bibfield  {journal} {\bibinfo
  {journal} {Phys. Rev. B}\ }\textbf {\bibinfo {volume} {85}},\ \bibinfo
  {pages} {064503} (\bibinfo {year} {2012})}\BibitemShut {NoStop}%
\bibitem [{\citenamefont {Cherpak}\ \emph {et~al.}(2014)\citenamefont
  {Cherpak}, \citenamefont {Lavrinovich}, \citenamefont {Gubin},\ and\
  \citenamefont {Vitusevich}}]{Che14apl}%
  \BibitemOpen
  \bibfield  {author} {\bibinfo {author} {\bibfnamefont {N.~T.}\ \bibnamefont
  {Cherpak}}, \bibinfo {author} {\bibfnamefont {A.~A.}\ \bibnamefont
  {Lavrinovich}}, \bibinfo {author} {\bibfnamefont {A.~I.}\ \bibnamefont
  {Gubin}}, \ and\ \bibinfo {author} {\bibfnamefont {S.~A.}\ \bibnamefont
  {Vitusevich}},\ }\href {\doibase 10.1063/1.4890123} {\bibfield  {journal}
  {\bibinfo  {journal} {Appl. Phys. Lett.}\ }\textbf {\bibinfo {volume}
  {105}},\ \bibinfo {pages} {022601} (\bibinfo {year} {2014})}\BibitemShut
  {NoStop}%
\bibitem [{\citenamefont {Lara}\ \emph {et~al.}(2015)\citenamefont {Lara},
  \citenamefont {Aliev}, \citenamefont {Silhanek},\ and\ \citenamefont
  {Moshchalkov}}]{Lar15nsr}%
  \BibitemOpen
  \bibfield  {author} {\bibinfo {author} {\bibfnamefont {A.}~\bibnamefont
  {Lara}}, \bibinfo {author} {\bibfnamefont {F.~G.}\ \bibnamefont {Aliev}},
  \bibinfo {author} {\bibfnamefont {A.~V.}\ \bibnamefont {Silhanek}}, \ and\
  \bibinfo {author} {\bibfnamefont {V.~V.}\ \bibnamefont {Moshchalkov}},\
  }\href {\doibase http://dx.doi.org/10.1038/srep09187} {\bibfield  {journal}
  {\bibinfo  {journal} {Sci. Rep.}\ }\textbf {\bibinfo {volume} {5}},\ \bibinfo
  {pages} {9187} (\bibinfo {year} {2015})}\BibitemShut {NoStop}%
\bibitem [{\citenamefont {Lara}\ \emph {et~al.}(2017)\citenamefont {Lara},
  \citenamefont {Aliev}, \citenamefont {Moshchalkov},\ and\ \citenamefont
  {Galperin}}]{Lar17pra}%
  \BibitemOpen
  \bibfield  {author} {\bibinfo {author} {\bibfnamefont {A.}~\bibnamefont
  {Lara}}, \bibinfo {author} {\bibfnamefont {F.~G.}\ \bibnamefont {Aliev}},
  \bibinfo {author} {\bibfnamefont {V.~V.}\ \bibnamefont {Moshchalkov}}, \ and\
  \bibinfo {author} {\bibfnamefont {Y.~M.}\ \bibnamefont {Galperin}},\ }\href
  {\doibase 10.1103/PhysRevApplied.8.034027} {\bibfield  {journal} {\bibinfo
  {journal} {Phys. Rev. Appl.}\ }\textbf {\bibinfo {volume} {8}},\ \bibinfo
  {pages} {034027} (\bibinfo {year} {2017})}\BibitemShut {NoStop}%
\bibitem [{\citenamefont {Madan}\ \emph {et~al.}(2018)\citenamefont {Madan},
  \citenamefont {Buh}, \citenamefont {Baranov}, \citenamefont {Kabanov},
  \citenamefont {Mrzel},\ and\ \citenamefont {Mihailovic}}]{Mad18sca}%
  \BibitemOpen
  \bibfield  {author} {\bibinfo {author} {\bibfnamefont {I.}~\bibnamefont
  {Madan}}, \bibinfo {author} {\bibfnamefont {J.}~\bibnamefont {Buh}}, \bibinfo
  {author} {\bibfnamefont {V.~V.}\ \bibnamefont {Baranov}}, \bibinfo {author}
  {\bibfnamefont {V.~V.}\ \bibnamefont {Kabanov}}, \bibinfo {author}
  {\bibfnamefont {A.}~\bibnamefont {Mrzel}}, \ and\ \bibinfo {author}
  {\bibfnamefont {D.}~\bibnamefont {Mihailovic}},\ }\href
  {http://advances.sciencemag.org/content/4/3/eaao0043} {\bibfield  {journal}
  {\bibinfo  {journal} {Sci. Adv.}\ }\textbf {\bibinfo {volume} {4}} (\bibinfo
  {year} {2018})}\BibitemShut {NoStop}%
\bibitem [{\citenamefont {Dobrovolskiy}\ \emph
  {et~al.}(2019{\natexlab{b}})\citenamefont {Dobrovolskiy}, \citenamefont
  {Bevz}, \citenamefont {Begun}, \citenamefont {Sachser}, \citenamefont
  {Vovk},\ and\ \citenamefont {Huth}}]{Dob19pra}%
  \BibitemOpen
  \bibfield  {author} {\bibinfo {author} {\bibfnamefont {O.~V.}\ \bibnamefont
  {Dobrovolskiy}}, \bibinfo {author} {\bibfnamefont {V.~M.}\ \bibnamefont
  {Bevz}}, \bibinfo {author} {\bibfnamefont {E.}~\bibnamefont {Begun}},
  \bibinfo {author} {\bibfnamefont {R.}~\bibnamefont {Sachser}}, \bibinfo
  {author} {\bibfnamefont {R.~V.}\ \bibnamefont {Vovk}}, \ and\ \bibinfo
  {author} {\bibfnamefont {M.}~\bibnamefont {Huth}},\ }\href {\doibase
  10.1103/PhysRevApplied.11.054064} {\bibfield  {journal} {\bibinfo  {journal}
  {Phys. Rev. Appl.}\ }\textbf {\bibinfo {volume} {11}},\ \bibinfo {pages}
  {054064} (\bibinfo {year} {2019}{\natexlab{b}})}\BibitemShut {NoStop}%
\bibitem [{\citenamefont {Bezuglyj}\ \emph {et~al.}(2019)\citenamefont
  {Bezuglyj}, \citenamefont {Shklovskij}, \citenamefont {Vovk}, \citenamefont
  {Bevz}, \citenamefont {Huth},\ and\ \citenamefont {Dobrovolskiy}}]{Bez19prb}%
  \BibitemOpen
  \bibfield  {author} {\bibinfo {author} {\bibfnamefont {A.~I.}\ \bibnamefont
  {Bezuglyj}}, \bibinfo {author} {\bibfnamefont {V.~A.}\ \bibnamefont
  {Shklovskij}}, \bibinfo {author} {\bibfnamefont {R.~V.}\ \bibnamefont
  {Vovk}}, \bibinfo {author} {\bibfnamefont {V.~M.}\ \bibnamefont {Bevz}},
  \bibinfo {author} {\bibfnamefont {M.}~\bibnamefont {Huth}}, \ and\ \bibinfo
  {author} {\bibfnamefont {O.~V.}\ \bibnamefont {Dobrovolskiy}},\ }\href
  {\doibase 10.1103/PhysRevB.99.174518} {\bibfield  {journal} {\bibinfo
  {journal} {Phys. Rev. B}\ }\textbf {\bibinfo {volume} {99}},\ \bibinfo
  {pages} {174518} (\bibinfo {year} {2019})}\BibitemShut {NoStop}%
\bibitem [{\citenamefont {Dobrovolskiy}\ \emph
  {et~al.}(2019{\natexlab{c}})\citenamefont {Dobrovolskiy}, \citenamefont
  {Sachser}, \citenamefont {Bevz}, \citenamefont {Lara}, \citenamefont {Aliev},
  \citenamefont {Shklovskij}, \citenamefont {Bezuglyj}, \citenamefont {Vovk},\
  and\ \citenamefont {Huth}}]{Dob19rrl}%
  \BibitemOpen
  \bibfield  {author} {\bibinfo {author} {\bibfnamefont {O.~V.}\ \bibnamefont
  {Dobrovolskiy}}, \bibinfo {author} {\bibfnamefont {R.}~\bibnamefont
  {Sachser}}, \bibinfo {author} {\bibfnamefont {V.~M.}\ \bibnamefont {Bevz}},
  \bibinfo {author} {\bibfnamefont {A.}~\bibnamefont {Lara}}, \bibinfo {author}
  {\bibfnamefont {F.~G.}\ \bibnamefont {Aliev}}, \bibinfo {author}
  {\bibfnamefont {V.~A.}\ \bibnamefont {Shklovskij}}, \bibinfo {author}
  {\bibfnamefont {A.}~\bibnamefont {Bezuglyj}}, \bibinfo {author}
  {\bibfnamefont {R.~V.}\ \bibnamefont {Vovk}}, \ and\ \bibinfo {author}
  {\bibfnamefont {M.}~\bibnamefont {Huth}},\ }\href {\doibase
  10.1002/pssr.201800223} {\bibfield  {journal} {\bibinfo  {journal} {Phys.
  Stat. Sol. - RRL}\ }\textbf {\bibinfo {volume} {13}},\ \bibinfo {pages}
  {1800223} (\bibinfo {year} {2019}{\natexlab{c}})}\BibitemShut {NoStop}%
\bibitem [{\citenamefont {Gorkov}\ and\ \citenamefont
  {Eliashberg}(1968)}]{Gor68etp}%
  \BibitemOpen
  \bibfield  {author} {\bibinfo {author} {\bibfnamefont {L.}~\bibnamefont
  {Gorkov}}\ and\ \bibinfo {author} {\bibfnamefont {G.~M.}\ \bibnamefont
  {Eliashberg}},\ }\href@noop {} {\bibfield  {journal} {\bibinfo  {journal}
  {Soviet Phys. JETP}\ }\textbf {\bibinfo {volume} {27}},\ \bibinfo {pages}
  {328} (\bibinfo {year} {1968})}\BibitemShut {NoStop}%
\bibitem [{\citenamefont {Aranson}\ and\ \citenamefont
  {Kramer}(2002)}]{Ara02rmp}%
  \BibitemOpen
  \bibfield  {author} {\bibinfo {author} {\bibfnamefont {I.~S.}\ \bibnamefont
  {Aranson}}\ and\ \bibinfo {author} {\bibfnamefont {L.}~\bibnamefont
  {Kramer}},\ }\href {\doibase 10.1103/RevModPhys.74.99} {\bibfield  {journal}
  {\bibinfo  {journal} {Rev. Mod. Phys.}\ }\textbf {\bibinfo {volume} {74}},\
  \bibinfo {pages} {99} (\bibinfo {year} {2002})}\BibitemShut {NoStop}%
\bibitem [{\citenamefont {Kwok}\ \emph {et~al.}(2016)\citenamefont {Kwok},
  \citenamefont {Welp}, \citenamefont {Glatz}, \citenamefont {Koshelev},
  \citenamefont {Kihlstrom},\ and\ \citenamefont {Crabtree}}]{Kwo16rpp}%
  \BibitemOpen
  \bibfield  {author} {\bibinfo {author} {\bibfnamefont {W.-K.}\ \bibnamefont
  {Kwok}}, \bibinfo {author} {\bibfnamefont {U.}~\bibnamefont {Welp}}, \bibinfo
  {author} {\bibfnamefont {A.}~\bibnamefont {Glatz}}, \bibinfo {author}
  {\bibfnamefont {A.~E.}\ \bibnamefont {Koshelev}}, \bibinfo {author}
  {\bibfnamefont {K.~J.}\ \bibnamefont {Kihlstrom}}, \ and\ \bibinfo {author}
  {\bibfnamefont {G.~W.}\ \bibnamefont {Crabtree}},\ }\href {\doibase
  10.1088/0034-4885/79/11/116501} {\bibfield  {journal} {\bibinfo  {journal}
  {Rep. Prog. Phys.}\ }\textbf {\bibinfo {volume} {79}},\ \bibinfo {pages}
  {116501} (\bibinfo {year} {2016})}\BibitemShut {NoStop}%
\bibitem [{\citenamefont {Hern\'andez}\ and\ \citenamefont
  {Dom\'{\i}nguez}(2008)}]{Her08prb}%
  \BibitemOpen
  \bibfield  {author} {\bibinfo {author} {\bibfnamefont {A.~D.}\ \bibnamefont
  {Hern\'andez}}\ and\ \bibinfo {author} {\bibfnamefont {D.}~\bibnamefont
  {Dom\'{\i}nguez}},\ }\href {\doibase 10.1103/PhysRevB.77.224505} {\bibfield
  {journal} {\bibinfo  {journal} {Phys. Rev. B}\ }\textbf {\bibinfo {volume}
  {77}},\ \bibinfo {pages} {224505} (\bibinfo {year} {2008})}\BibitemShut
  {NoStop}%
\bibitem [{\citenamefont {Oripov}\ and\ \citenamefont {Anlage}()}]{Ori19arx}%
  \BibitemOpen
  \bibfield  {author} {\bibinfo {author} {\bibfnamefont {B.}~\bibnamefont
  {Oripov}}\ and\ \bibinfo {author} {\bibfnamefont {S.~M.}\ \bibnamefont
  {Anlage}},\ }\href@noop {} {\bibinfo  {journal} {arXiv:1909.02714}\
  }\BibitemShut {NoStop}%
\bibitem [{\citenamefont {Gropp}\ \emph {et~al.}(1996)\citenamefont {Gropp},
  \citenamefont {Kaper}, \citenamefont {Leaf}, \citenamefont {Levine},
  \citenamefont {Palumbo},\ and\ \citenamefont {Vinokur}}]{Gro96jcp}%
  \BibitemOpen
\bibfield  {journal} {  }\bibfield  {author} {\bibinfo {author} {\bibfnamefont
  {W.~D.}\ \bibnamefont {Gropp}}, \bibinfo {author} {\bibfnamefont {H.~G.}\
  \bibnamefont {Kaper}}, \bibinfo {author} {\bibfnamefont {G.~K.}\ \bibnamefont
  {Leaf}}, \bibinfo {author} {\bibfnamefont {D.~M.}\ \bibnamefont {Levine}},
  \bibinfo {author} {\bibfnamefont {M.}~\bibnamefont {Palumbo}}, \ and\
  \bibinfo {author} {\bibfnamefont {V.~M.}\ \bibnamefont {Vinokur}},\ }\href
  {http://www.sciencedirect.com/science/article/pii/S0021999196900224}
  {\bibfield  {journal} {\bibinfo  {journal} {J. Comput. Phys.}\ }\textbf
  {\bibinfo {volume} {123}},\ \bibinfo {pages} {254} (\bibinfo {year}
  {1996})}\BibitemShut {NoStop}%
\bibitem [{\citenamefont {Kato}\ \emph {et~al.}(1993)\citenamefont {Kato},
  \citenamefont {Enomoto},\ and\ \citenamefont {Maekawa}}]{Kat93prb}%
  \BibitemOpen
  \bibfield  {author} {\bibinfo {author} {\bibfnamefont {R.}~\bibnamefont
  {Kato}}, \bibinfo {author} {\bibfnamefont {Y.}~\bibnamefont {Enomoto}}, \
  and\ \bibinfo {author} {\bibfnamefont {S.}~\bibnamefont {Maekawa}},\ }\href
  {\doibase 10.1103/PhysRevB.47.8016} {\bibfield  {journal} {\bibinfo
  {journal} {Phys. Rev. B}\ }\textbf {\bibinfo {volume} {47}},\ \bibinfo
  {pages} {8016} (\bibinfo {year} {1993})}\BibitemShut {NoStop}%
\bibitem [{\citenamefont {Brandt}(1995)}]{Bra95rpp}%
  \BibitemOpen
  \bibfield  {author} {\bibinfo {author} {\bibfnamefont {E.~H.}\ \bibnamefont
  {Brandt}},\ }\href {http://stacks.iop.org/0034-4885/58/i=11/a=003} {\bibfield
   {journal} {\bibinfo  {journal} {Rep. Progr. Phys.}\ }\textbf {\bibinfo
  {volume} {58}},\ \bibinfo {pages} {1465} (\bibinfo {year}
  {1995})}\BibitemShut {NoStop}%
\bibitem [{\citenamefont {Anthore}\ \emph {et~al.}(2003)\citenamefont
  {Anthore}, \citenamefont {Pothier},\ and\ \citenamefont {Esteve}}]{Ant03prl}%
  \BibitemOpen
  \bibfield  {author} {\bibinfo {author} {\bibfnamefont {A.}~\bibnamefont
  {Anthore}}, \bibinfo {author} {\bibfnamefont {H.}~\bibnamefont {Pothier}}, \
  and\ \bibinfo {author} {\bibfnamefont {D.}~\bibnamefont {Esteve}},\ }\href
  {\doibase 10.1103/PhysRevLett.90.127001} {\bibfield  {journal} {\bibinfo
  {journal} {Phys. Rev. Lett.}\ }\textbf {\bibinfo {volume} {90}},\ \bibinfo
  {pages} {127001} (\bibinfo {year} {2003})}\BibitemShut {NoStop}%
\bibitem [{\citenamefont {Anderson}(1958)}]{And58prv}%
  \BibitemOpen
  \bibfield  {author} {\bibinfo {author} {\bibfnamefont {P.~W.}\ \bibnamefont
  {Anderson}},\ }\href {\doibase 10.1103/PhysRev.110.827} {\bibfield  {journal}
  {\bibinfo  {journal} {Phys. Rev.}\ }\textbf {\bibinfo {volume} {110}},\
  \bibinfo {pages} {827} (\bibinfo {year} {1958})}\BibitemShut {NoStop}%
\bibitem [{\citenamefont {Maki}\ and\ \citenamefont {Fulde}(1965)}]{Mak65prv}%
  \BibitemOpen
  \bibfield  {author} {\bibinfo {author} {\bibfnamefont {K.}~\bibnamefont
  {Maki}}\ and\ \bibinfo {author} {\bibfnamefont {P.}~\bibnamefont {Fulde}},\
  }\href {\doibase 10.1103/PhysRev.140.A1586} {\bibfield  {journal} {\bibinfo
  {journal} {Phys. Rev.}\ }\textbf {\bibinfo {volume} {140}},\ \bibinfo {pages}
  {A1586} (\bibinfo {year} {1965})}\BibitemShut {NoStop}%
\bibitem [{\citenamefont {Mooij}(1981)}]{Moo81inb}%
  \BibitemOpen
  \bibfield  {author} {\bibinfo {author} {\bibfnamefont {J.~E.}\ \bibnamefont
  {Mooij}},\ }\enquote {\bibinfo {title} {Enhancement of superconductivity},}\
  \ (\bibinfo  {publisher} {Plenum Press, New York},\ \bibinfo {year} {1981})\
  Chap.~\bibinfo {chapter} {9}, pp.\ \bibinfo {pages} {191--287}\BibitemShut
  {NoStop}%
\bibitem [{\citenamefont {Larkin}\ and\ \citenamefont
  {Ovchinnikov}(1976)}]{Lar76etp}%
  \BibitemOpen
  \bibfield  {author} {\bibinfo {author} {\bibfnamefont {A.~I.}\ \bibnamefont
  {Larkin}}\ and\ \bibinfo {author} {\bibfnamefont {Y.~N.}\ \bibnamefont
  {Ovchinnikov}},\ }\href@noop {} {\bibfield  {journal} {\bibinfo  {journal}
  {Sov. Phys. JETP}\ }\textbf {\bibinfo {volume} {41}},\ \bibinfo {pages} {960}
  (\bibinfo {year} {1976})}\BibitemShut {NoStop}%
\bibitem [{\citenamefont {Aslamazov}\ and\ \citenamefont
  {Lempitskii}(1982)}]{Asl82etp}%
  \BibitemOpen
  \bibfield  {author} {\bibinfo {author} {\bibfnamefont {L.~G.}\ \bibnamefont
  {Aslamazov}}\ and\ \bibinfo {author} {\bibfnamefont {S.~V.}\ \bibnamefont
  {Lempitskii}},\ }\href@noop {} {\bibfield  {journal} {\bibinfo  {journal}
  {Zh. Eksp. Teor. Fiz.}\ }\textbf {\bibinfo {volume} {82}},\ \bibinfo {pages}
  {1671} (\bibinfo {year} {1982})}\BibitemShut {NoStop}%
\bibitem [{\citenamefont {Lefloch}\ \emph {et~al.}(1999)\citenamefont
  {Lefloch}, \citenamefont {Hoffmann},\ and\ \citenamefont
  {Demolliens}}]{Lef99pcs}%
  \BibitemOpen
  \bibfield  {author} {\bibinfo {author} {\bibfnamefont {F.}~\bibnamefont
  {Lefloch}}, \bibinfo {author} {\bibfnamefont {C.}~\bibnamefont {Hoffmann}}, \
  and\ \bibinfo {author} {\bibfnamefont {O.}~\bibnamefont {Demolliens}},\
  }\href {http://www.sciencedirect.com/science/article/pii/S092145349900297X}
  {\bibfield  {journal} {\bibinfo  {journal} {Physica C}\ }\textbf {\bibinfo
  {volume} {319}},\ \bibinfo {pages} {258} (\bibinfo {year}
  {1999})}\BibitemShut {NoStop}%
\bibitem [{\citenamefont {Kaplan}\ \emph {et~al.}(1976)\citenamefont {Kaplan},
  \citenamefont {Chi}, \citenamefont {Langenberg}, \citenamefont {Chang},
  \citenamefont {Jafarey},\ and\ \citenamefont {Scalapino}}]{Kap76prb}%
  \BibitemOpen
  \bibfield  {author} {\bibinfo {author} {\bibfnamefont {S.~B.}\ \bibnamefont
  {Kaplan}}, \bibinfo {author} {\bibfnamefont {C.~C.}\ \bibnamefont {Chi}},
  \bibinfo {author} {\bibfnamefont {D.~N.}\ \bibnamefont {Langenberg}},
  \bibinfo {author} {\bibfnamefont {J.~J.}\ \bibnamefont {Chang}}, \bibinfo
  {author} {\bibfnamefont {S.}~\bibnamefont {Jafarey}}, \ and\ \bibinfo
  {author} {\bibfnamefont {D.~J.}\ \bibnamefont {Scalapino}},\ }\href {\doibase
  10.1103/PhysRevB.14.4854} {\bibfield  {journal} {\bibinfo  {journal} {Phys.
  Rev. B}\ }\textbf {\bibinfo {volume} {14}},\ \bibinfo {pages} {4854}
  (\bibinfo {year} {1976})}\BibitemShut {NoStop}%
\bibitem [{\citenamefont {Schmid}(1968)}]{Sch68pkm}%
  \BibitemOpen
  \bibfield  {author} {\bibinfo {author} {\bibfnamefont {A.}~\bibnamefont
  {Schmid}},\ }\href {\doibase 10.1007/BF02422735} {\bibfield  {journal}
  {\bibinfo  {journal} {Phys. Kond. Mater.}\ }\textbf {\bibinfo {volume} {8}},\
  \bibinfo {pages} {129} (\bibinfo {year} {1968})}\BibitemShut {NoStop}%
\bibitem [{\citenamefont {Tinkham}(2004)}]{Tin04boo}%
  \BibitemOpen
  \bibfield  {author} {\bibinfo {author} {\bibfnamefont {M.}~\bibnamefont
  {Tinkham}},\ }\href@noop {} {\emph {\bibinfo {title} {Introduction to
  Superconductivity}}}\ (\bibinfo  {publisher} {Mineola, New York},\ \bibinfo
  {year} {2004})\BibitemShut {NoStop}%
\bibitem [{\citenamefont {Dobrovolskiy}\ and\ \citenamefont
  {Huth}(2012)}]{Dob12tsf}%
  \BibitemOpen
  \bibfield  {author} {\bibinfo {author} {\bibfnamefont {O.~V.}\ \bibnamefont
  {Dobrovolskiy}}\ and\ \bibinfo {author} {\bibfnamefont {M.}~\bibnamefont
  {Huth}},\ }\href {\doibase 10.1016/j.tsf.2012.04.083} {\bibfield  {journal}
  {\bibinfo  {journal} {Thin Solid Films}\ }\textbf {\bibinfo {volume} {520}},\
  \bibinfo {pages} {5985} (\bibinfo {year} {2012})}\BibitemShut {NoStop}%
\bibitem [{\citenamefont {Gubin}\ \emph {et~al.}(2005)\citenamefont {Gubin},
  \citenamefont {Il'in}, \citenamefont {Vitusevich}, \citenamefont {Siegel},\
  and\ \citenamefont {Klein}}]{Gub05prb}%
  \BibitemOpen
  \bibfield  {author} {\bibinfo {author} {\bibfnamefont {A.~I.}\ \bibnamefont
  {Gubin}}, \bibinfo {author} {\bibfnamefont {K.~S.}\ \bibnamefont {Il'in}},
  \bibinfo {author} {\bibfnamefont {S.~A.}\ \bibnamefont {Vitusevich}},
  \bibinfo {author} {\bibfnamefont {M.}~\bibnamefont {Siegel}}, \ and\ \bibinfo
  {author} {\bibfnamefont {N.}~\bibnamefont {Klein}},\ }\href {\doibase
  10.1103/PhysRevB.72.064503} {\bibfield  {journal} {\bibinfo  {journal} {Phys.
  Rev. B}\ }\textbf {\bibinfo {volume} {72}},\ \bibinfo {pages} {064503}
  (\bibinfo {year} {2005})}\BibitemShut {NoStop}%
\bibitem [{\citenamefont {Lehnert}\ \emph {et~al.}(1994)\citenamefont
  {Lehnert}, \citenamefont {Schuster},\ and\ \citenamefont
  {Gundlach}}]{Leh94apl}%
  \BibitemOpen
  \bibfield  {author} {\bibinfo {author} {\bibfnamefont {T.}~\bibnamefont
  {Lehnert}}, \bibinfo {author} {\bibfnamefont {K.}~\bibnamefont {Schuster}}, \
  and\ \bibinfo {author} {\bibfnamefont {K.~H.}\ \bibnamefont {Gundlach}},\
  }\href {\doibase 10.1063/1.113051} {\bibfield  {journal} {\bibinfo  {journal}
  {Appl. Phys. Lett.}\ }\textbf {\bibinfo {volume} {65}},\ \bibinfo {pages}
  {112} (\bibinfo {year} {1994})}\BibitemShut {NoStop}%
\bibitem [{\citenamefont {Pronin}\ \emph {et~al.}(1998)\citenamefont {Pronin},
  \citenamefont {Dressel}, \citenamefont {Pimenov}, \citenamefont {Loidl},
  \citenamefont {Roshchin},\ and\ \citenamefont {Greene}}]{Pro98prb}%
  \BibitemOpen
  \bibfield  {author} {\bibinfo {author} {\bibfnamefont {A.~V.}\ \bibnamefont
  {Pronin}}, \bibinfo {author} {\bibfnamefont {M.}~\bibnamefont {Dressel}},
  \bibinfo {author} {\bibfnamefont {A.}~\bibnamefont {Pimenov}}, \bibinfo
  {author} {\bibfnamefont {A.}~\bibnamefont {Loidl}}, \bibinfo {author}
  {\bibfnamefont {I.~V.}\ \bibnamefont {Roshchin}}, \ and\ \bibinfo {author}
  {\bibfnamefont {L.~H.}\ \bibnamefont {Greene}},\ }\href {\doibase
  10.1103/PhysRevB.57.14416} {\bibfield  {journal} {\bibinfo  {journal} {Phys.
  Rev. B}\ }\textbf {\bibinfo {volume} {57}},\ \bibinfo {pages} {14416}
  (\bibinfo {year} {1998})}\BibitemShut {NoStop}%
\bibitem [{\citenamefont {Buscaglia}\ \emph {et~al.}(2000)\citenamefont
  {Buscaglia}, \citenamefont {Bolech},\ and\ \citenamefont
  {L{\'o}pez}}]{Bus00inb}%
  \BibitemOpen
  \bibfield  {author} {\bibinfo {author} {\bibfnamefont {G.~C.}\ \bibnamefont
  {Buscaglia}}, \bibinfo {author} {\bibfnamefont {C.}~\bibnamefont {Bolech}}, \
  and\ \bibinfo {author} {\bibfnamefont {A.}~\bibnamefont {L{\'o}pez}},\
  }\enquote {\bibinfo {title} {Connectivity and superconductivity},}\ \
  (\bibinfo  {publisher} {Springer Berlin Heidelberg},\ \bibinfo {address}
  {Berlin, Heidelberg},\ \bibinfo {year} {2000})\ Chap.\ \bibinfo {chapter} {On
  the Numerical Solution of the Time-Dependent {Ginzburg-Landau} Equations in
  Multiply Connected Domains}, pp.\ \bibinfo {pages} {200--214}\BibitemShut
  {NoStop}%
\bibitem [{\citenamefont {Lara}\ \emph {et~al.}(2020)\citenamefont {Lara},
  \citenamefont {Gonzalez-Ruano},\ and\ \citenamefont {Aliev}}]{Lar20ltp}%
  \BibitemOpen
  \bibfield  {author} {\bibinfo {author} {\bibfnamefont {A.}~\bibnamefont
  {Lara}}, \bibinfo {author} {\bibfnamefont {C.}~\bibnamefont
  {Gonzalez-Ruano}}, \ and\ \bibinfo {author} {\bibfnamefont {F.~G.}\
  \bibnamefont {Aliev}},\ }\href {https://arxiv.org/abs/2001.07971} {\bibfield
  {journal} {\bibinfo  {journal} {Low Temp. Phys./Fiz. Nizk. Temp.}\ }\textbf
  {\bibinfo {volume} {46}},\ \bibinfo {pages} {386} (\bibinfo {year}
  {2020})}\BibitemShut {NoStop}%
\end{thebibliography}
%

\end{document}